\documentclass{article}
\usepackage[letterpaper, portrait, margin=1in]{geometry}
\usepackage[english]{babel}
\usepackage[utf8]{inputenc}
\usepackage[T1]{fontenc}
\usepackage[superscript]{cite}
\usepackage{lmodern}
\usepackage{helvet, etoolbox, titlesec, caption, booktabs, xcolor, caption, float, scrextend, fancyhdr, amsmath, amssymb, bm, bbm, dsfont, graphicx, setspace}
\graphicspath{{images/}}
\newcommand{\E}{\mathbb{E}}

\newcommand{\Xb}{\bm{X}}

\newcommand{\Zb}{\bm{Z}}

\newcommand{\bgamma}{\bm{\gamma}}

\newcommand{\bbeta}{\bm{\beta}}
\newcommand{\beeta}{\bm{\eta}}
\newcommand{\dd}{\mathrm{d}}
\newcommand{\p}{\mathrm{P}}
\newcommand{\Hcal}{\mathcal{H}}
\newcommand{\1}{\mathds{1}}
\fancypagestyle{plain}{
	\fancyhf{}

}

\makeatletter
\patchcmd{\@maketitle}{\LARGE \@title}{\fontsize{16}{19.2}\selectfont\@title}{}{}
\makeatother

\usepackage{authblk}

\newsavebox\affbox
\author[1,2*]{\textbf{George Stefan}}
\author[1,2]{\textbf{Eleanor Pullenayegum}}
\affil[1]{Dalla Lana School of Public Health, University of Toronto, Ontario, Canada}
\affil[2]{Child Health Evaluative Sciences, The Hospital for Sick Children, Ontario, Canada}

\titlespacing\section{0pt}{12pt plus 4pt minus 2pt}{0pt plus 2pt minus 2pt}
\titlespacing\subsection{12pt}{12pt plus 4pt minus 2pt}{0pt plus 2pt minus 2pt}
\titlespacing\subsubsection{12pt}{12pt plus 4pt minus 2pt}{0pt plus 2pt minus 2pt}

\titleformat{\section}{\normalfont\fontsize{10}{15}\bfseries}{\thesection.}{1em}{}
\titleformat{\subsection}{\normalfont\fontsize{10}{15}\bfseries}{\thesubsection.}{1em}{}
\titleformat{\subsubsection}{\normalfont\fontsize{10}{15}\bfseries}{\thesubsubsection.}{1em}{}

\titleformat{\author}{\normalfont\fontsize{10}{15}\bfseries}{\thesection}{1em}{}

\title{\textbf{\huge Inverse-intensity weighted generalized estimating equations with irregularly measured longitudinal data and informative dropout}}
\date{}    

\begin{document}

\pagestyle{headings}	
\newpage
\setcounter{page}{1}
\renewcommand{\thepage}{\arabic{page}}
	
\captionsetup[figure]{labelfont={bf},labelformat={default},labelsep=period,name={Figure }}	\captionsetup[table]{labelfont={bf},labelformat={default},labelsep=period,name={Table }}
\setlength{\parskip}{0.5em}
	
\maketitle
	
\noindent\rule{15.9cm}{0.5pt}
	\begin{abstract}
Longitudinal data are commonly encountered in biomedical research, including randomized trials and retrospective cohort studies. Subjects are typically followed over a period of time and may be scheduled for follow-up at pre-determined time points. However, subjects may miss their appointments or return at non-specified times, leading to irregularity in the visit process. IIW-GEEs have been developed as one method to account for this irregularity, whereby estimates from a visit intensity model are used as weights in a GEE model with an independent correlation structure. We show that currently available methods can be biased for situations in which the health outcome of interest may influence a subject’s dropout from the study. We have extended the IIW-GEE framework to adjust for informative dropout and have demonstrated via simulation studies that this bias can be significantly reduced. We have illustrated this method using the STAR*D clinical trial data, and observed that the disease trajectory was generally overestimated when informative dropout was not accounted for.
		\let\thefootnote\relax\footnotetext{
			\small $^{*}$\textbf{George Stefan.} \textit{
				\textit{E-mail address: \color{cyan}george.stefan@mail.utoronto.ca}}}
	\end{abstract}

\newpage

\section{Introduction}

Longitudinal data are commonly used to estimate disease trajectories as a function of potential prognostic factors. However, both observation times and dropout may be associated with the health outcome, which if left unaddressed may introduce bias in the estimation. Patients receiving treatment may be scheduled for follow-up at predetermined time intervals; however, they may miss their appointments or return at unspecified times, leading to irregular measurements. This is of particular importance when the timing of the visits is associated with the course of the disease; for example, a patient may experience a flare and return for follow-up earlier than the schedule prescribes or may visit more frequently during periods of high disease activity. A patient may also drop out of a study, whereupon their disease trajectory would no longer be observed. In certain cases, the dropout may be informative in the sense that it may be influenced by the progression of the disease. We will propose methodology for longitudinal data exhibiting outcome-dependent follow-up which addresses potential problems due to irregularity and informative dropout by combining inverse-intensity of visit and inverse-probability of dropout weights with generalized estimating equations.

In the context of regularly spaced longitudinal data subject to dropout, Robins et al.\cite{robins1995analysis} first proposed incorporating inverse-probability weights (IPW) of being observed into the generalized estimating equation (GEE) structure. These weights could be estimated via some dropout model, e.g. logistic regression, and provided that the dropout model is correctly specified and the data is missing at random, the weighted GEE estimates are consistent.\cite{seaman2009doubly} In the context of marginal structural models, Cole and Hern\'{a}n\cite{cole2008constructing} multiplied inverse-probabilities of exposure by inverse-probabilities of censoring to obtain weights which allow the model to represent a population with neither confounding nor dropout.

Approaches to handling irregularly observed longitudinal data include inverse-intensity weighted generalized estimating equations (IIW-GEEs)\cite{lin2004analysis, buzkova2007longitudinal}; Lin-Ying models which employ estimating equations incorporating a quasi-residual term containing the intensity model estimates\cite{lin2001semiparametric, lin2003semiparametric, buzkova2009semiparametric}; and semiparametric joint models \cite{sun2011semiparametric, miao2016analyzing, he2016joint, su2018semiparametric, shen2019regression, han2014joint, yu2019additive}. Lin and Ying\cite{lin2003semiparametric} deal with informative dropout via artificial censoring, whereby the informative dropout times are assumed to follow an accelerated failure time model, which is used to to attenuate the observed dropout times. Miao et al.\cite{miao2016analyzing}, He et al.\cite{he2016joint} and Shen et al.\cite{shen2019regression} have jointly addressed outcome-dependent follow-up and informative dropout by incorporating latent variables shared by the outcome and intensity models, while Han et al.\cite{han2014joint} and Yu et al.\cite{yu2019additive} did so by using artificial censoring in a similar fashion to Lin and Ying \cite{lin2003semiparametric}. 

The methods which are able to deal with both outcome-dependent follow-up and informative dropout\cite{miao2016analyzing, he2016joint, su2018semiparametric, shen2019regression, han2014joint, yu2019additive} assume either that the same set of covariates is associated with the outcome, observation, and dropout processes, or that the visit intensity covariates are a subset of the outcome model covariates. We propose extending IIW-GEEs to account for informative dropout, while loosening these covariate restrictions. Neither the IIW-GEEs developed by Lin et al. \cite{lin2004analysis} nor the extensions by B\r{u}\v{z}kov\'{a} and Lumley \cite{buzkova2007longitudinal} account for informative dropout when estimating the marginal mean model. B\r{u}\v{z}kov\'{a} and Lumley \cite{buzkova2009semiparametric} expand on the work by Lin and Ying \cite{lin2001semiparametric} to allow for outcome-dependent follow-up, but do not incorporate informative dropout as in Lin and Ying \cite{lin2003semiparametric}.  

In a causal inference context, Cole and Hern\'{a}n\cite{cole2008constructing} multiplied two sets of inverse-probability weights to adjust for measured confounding and selection bias simultaneously; since then, Coulombe et al. \cite{coulombe2024multiply} have developed a doubly augmented inverse probability of treatment and intensity weighted estimator and provided a framework to accommodate informative censoring by including the censoring weights in the treatment model; however, the underlying theory is not presented in detail and there are no simulation results. Tompkins et al. \cite{tompkins2024flexible} suggested a similar approach of modifying a causal estimator which already multiplies inverse-intensity and inverse-treatment weights by further multiplying by censoring weights. Neither authors present the underlying theory in detail; Tompkins et al. present a simulation scenario in which they assess the impact of informative censoring, but only modelling a causal contrast without any time-dependent covariates. While multiplying by inverse-censoring weights may seem like a natural extension, we show the theory as it pertains to IIW-GEEs and suggest how to deal with complications arising in particular situations---for instance, when dropout from the study can only occur at a visit time.

We first introduce the notation, assumptions, models, and estimating equation. We then assess performance in terms of bias and variability under two simulation scenarios and illustrate our method using the STAR*D clinical trial data. Finally, we discuss potential difficulties with our approach and plans for future work.

\section{Methods}

\subsection{Notation and Models}

Visits occur at times $T_{i1}<T_{i2}<\cdots<T_{iK_i}$, where $K_i$ indicates the total number of visits for subject $i$. We count the total number of visits for subject $i$ up to time $t$ via $N_i(t)=\sum_{k=1}^{K_i} \1(T_{ik}\leq t)$ and denote $\dd N_i(t) = \1(\text{subject $i$ visits at time $t$})$, where $\1(\cdot)$ is the indicator function. We define $D_i$ as a time to informative dropout, in that a subject's dropout may be influenced by the outcome. We further define $G_i$ as a noninformative censoring time, and $L_i$ as a time for a competing event, which would preclude the outcome from continuing to be measured. We assume that the dropout event does not pose a competing risk---that is, the outcome of interest can still exist for any individual after their dropout event---and if a competing event exists, it is distinct from the dropout event.

We indicate via $\xi_i^D(t)=\1(D_i\leq t)$ that subject $i$ has informatively dropped out; via $\xi_i^G(t)=\1(G_i\leq t)$ that they have been noninformatively censored; via $\xi_i^L(t)=\1(L_i< t)$ that they have experienced a competing event before time $t$. We further indicate via $\zeta_i(t)=\1(D_i> t)\1(G_i> t)\1(L_i\geq t)$ that subject $i$ is still under follow-up and is competing event-free by time $t$. The observed visit process $N_i(t)$ can then be related to an underlying counterfactual process $N_i^*(t)$ via $N_i(t)=N_i^*(t \land C_i)$, where $C_i=\min(D_i,G_i,L_i)$. We denote the entire observed data history for a subject up to time $t$ by $\Hcal_i^O(t)=\{\bar{N}_i(t), \bar\xi_i^D(t\land C_i),\bar\xi_i^G(t\land C_i), \bar\xi_i^L(t\land C_i),\bar{\Xb}_i^O(t), \bar{\Zb}_i^O(t), \bar{Y}_i^O(t)\}$, where for example, $\bar{\Xb}_i(t)=\{\Xb_i(s): 0\leq s\leq t\}$ and $\bar{\Xb}_i^{O}(t)=\{\Xb_i(s): 0\leq s\leq t, \dd N_i(s)=1\}$. We define $\Zb_i(t)$ as a vector of auxiliary covariates, in that they may have an effect on the visit process but not directly on the outcome. The relationship between $Y_i(t)$ and possibly time-dependent covariates $ \Xb_i(t)$ is modelled via marginal mean model 
\begin{align*}
    g(\mu_i(t))=\bbeta_0^\top \Xb_i(t),
\end{align*}
where $\bbeta_0$ is a vector of parameters and we define $\mu_i(t)\equiv\E[Y_i(t)|\Xb_i(t), \bar\xi^L_i(t)=0]$, with $g(\cdot)$ as a monotonic and differentiable link function. Note that we have defined the outcome while conditioning on there not having been a competing event up to just before time $t$. We assume that this competing event time is predictable at time $t$---that is, knowing the history of the event up to just before $t$ allows us to fully determine the status at $t$. For the uncensored visit times and in the absence of a competing risk, we may adopt the proportional intensity model
\begin{align*}
    \E[\dd N_i^*(t)| \Hcal_i^O(t^-), \Xb_i(t)]=e^{\bgamma_0^\top\Hcal_i^O(t^-)}\dd \Lambda_0(t), 
\end{align*}
which induces the censored visit times model  
\begin{align}
    \E[\dd N_i(t) |\Hcal_i^O(t^-),\Xb_i(t)] &= \1(L_i\geq t)e^{\bgamma_0^\top\Hcal_i^O(t^-)}\dd\Lambda_0(t) \nonumber\\
    &\qquad \times \p(D_i> t|G_i> t, \Hcal_i^O(t^-),\Xb_i(t))\p(G_i> t|\Hcal_i^O(t^-),\Xb_i(t)),\label{fullbreakdown}
\end{align}
derived in Appendix 6.4. While the counterfactual visit times only depend on the history up to just before $t$ and the model is defined similarly to Lin et al. \cite{lin2004analysis} and B\r{u}\v{z}kov\'{a} and Lumley \cite{buzkova2007longitudinal}, our model for the observed visits differs by incorporating potentially informative dropout. We assume that noninformative censoring status after time $t$ is conditionally independent of $Y_i(t)$ given $\Xb_i(t)$, i.e.
\begin{align}
    \p[G_i>t|Y_i(t),\Xb_i(t)]=\p[G_i>t|\Xb_i(t)]. \label{assG}
\end{align}
For the visit process, we assume that a counterfactual visit at time $t$ is conditionally independent of follow-up status at time $t$, given the data history until just before time $t$, i.e.
\begin{align}
    \E[\dd N_i^*(t)| \zeta_i(t), \Hcal_i^O(t^-)] = \E[\dd N_i^*(t)|\Hcal_i^O(t^-)], \label{assdN}
\end{align}
Condition (\ref{assdN}) enables us to simplify the observed visit process model as in equation (\ref{fullbreakdown}). Condition (\ref{assG}) ensures that we do not need to explicitly model the noninformative censoring mechanism and leads to an unbiased estimating equation. 

\subsection{Estimation}
In order to estimate the parameter of interest $\bbeta_0$ under the GEE framework, we incorporate both the visit intensity and the probability of not having dropped out by a certain time point for each individual into the weights. The estimating function is given by
\begin{align}
    \bm{U}(\bbeta;\hat{\bgamma},\hat{\beeta},h)&=\sum_{i=1}^n \int_0^\infty \Xb_i(t)\left\{\frac{\dd g(\mu)}{\dd \mu}\biggr|_{\mu_i(t;\bbeta)} \right\}^{-1} v(\mu_i(t;\bbeta))^{-1}\times  \frac{Y_i(t)-\mu_i(t;\bbeta)}{\rho_i(t;\hat{\bgamma},\hat{\beeta},h)}\dd N_i(t), \label{GEE}
\end{align}
where $v(\mu_i(t;\bbeta))$ is a matrix representing the variance-covariance structure as a function of the mean. Instead of only using estimated rates from the visit model in the weights as in Lin et al. \cite{lin2004analysis} and B\r{u}\v{z}kov\'{a} and Lumley \cite{buzkova2007longitudinal}, we further multiply these by the inverse-probabilities of not yet having informatively dropped out at a given time, yielding
\begin{align}
    \rho_i(t;\bgamma_0,\beeta_0,h)=\frac{e^{\bgamma_0^\top \Hcal_i^O
    (t^-)}\times \p(D_i> t|G_i> t, \Hcal_i^O(t^-),\Xb_i(t); \beeta_0)}{h(\Xb_i(t))},\label{weights}
\end{align}
where $h(\cdot)$ is a stabilizing function. In order to solve (\ref{GEE}), we first obtain an estimate $\hat\bgamma$ of $\bgamma_0$ via the partial likelihood \cite{andersen2012statistical}. Equation (\ref{fullbreakdown}) implies that along with the typical intensity model, we must specify a conditional model for the probability that the subjects have not yet informatively dropped out by time $t$, allowing for dependency on the covariate history, which crucially may include the subject's outcome measured before time $t$. We made the dependency on some parameter $\beeta_0$ explicit in (\ref{GEE}) and (\ref{weights}), since we later use logistic regression to obtain estimate $\hat\beeta$ of $\beeta_0$ in our simulations and data example; however, the dropout model need not be parametric. We have shown that the estimating function has mean zero (Appendix 6.5) subject to a positivity assumption, and thus, by the theory of M-estimation, this will yield consistent and asymptotically unbiased estimates of $\bbeta_0$ under typical regularity conditions.

Equations (\ref{GEE}) and (\ref{weights}) are valid only if the informative dropout time $D_i$ occurs in continuous time; special consideration is needed for the case in which it can only occur at a visit time. For example, a patient may recover sufficiently that they are informed during a visit that they no longer need to return for future follow-up. In this situation, we cannot include $\bar\xi_i^D(t)$ in our history; the dropout status between the last observed visit and time $t$ is known and thus cannot be conditioned on. We re-define the history as $\mathcal{H}_i^{O\backslash D}(t)=\{\bar{N}_i(t),\bar\xi_i^G(t\land C_i), \bar\xi_i^L(t\land C_i),\bar{\Xb}_i^O(t), \bar{\Zb}_i^O(t), \bar{Y}_i^O(t)\}$ and note that equation (\ref{fullbreakdown}) still holds with this reduced history and the derivation in Appendix 6.5 follows similarly. The term related to the informative dropout could be further written as
\begin{align}
   &\p(D_i> t|G_i> t, \Hcal_i^{O\backslash D}(t^-))=\p(D_i> t| G_i> t, \Hcal_i^{O\backslash D}(T_{iN_i(t^-)}), N_i(t)=N_i(T_{iN_i(t^-)})) \nonumber \\
    &= \frac{\p(D_i>t, G_i> t, N_i(t) = N_i(T_{iN_i(t^-)})| \Hcal_i^{O\backslash D}(T_{iN_i(t^-)}))}{\p(N_i(t) = N_i(T_{iN_i(t^-)}), G_i> t| \Hcal_i^{O\backslash D}(T_{iN_i(t^-)}))} \nonumber\\
    &= \frac{\p(D_i> t| G_i> t, \Hcal_i^{O\backslash D}(T_{iN_i(t^-)}))\p(N_i(t) = N_i(T_{iN_i(t^-)})| D_i> t, G_i> t, \Hcal_i^{O\backslash D}(T_{iN_i(t^-)}))}{\p(N_i(t) = N_i(T_{iN_i(t^-)})|  G_i> t, \Hcal_i^{O\backslash D}(T_{iN_i(t^-)}))} \nonumber\\
    &= \frac{ \exp\left\{-\int_{T_{iN_i(t^-)}}^t \lambda_0(s)e^{\bgamma_0^\top\Hcal_i^{O\backslash D}(s)}\dd s\right\}}{\exp\left\{-\int_{T_{iN_i(t^-)}}^t \lambda_0(s)e^{\bgamma_0^\top\Hcal_i^{O\backslash D}(s)}\dd s\right\} + \text{odds of informative dropout at $T_{iN_i(t^-)}$}} \label{dropoutexplicit},
\end{align}
Lin et al. \cite{lin2004analysis} and B\r{u}\v{z}kov\'{a} and Lumley \cite{buzkova2007longitudinal} propose asymptotic standard errors for $\bbeta_0$ which account for the uncertainty in the estimation of intensity model parameter $\bgamma_0$, though these often undercover in small samples. They assume noninformative dropout, so they do not specify a model for the dropout, nor do they account for the associated uncertainty. In our setting, standard errors accounting for uncertainty in both the estimation of $\bgamma_0$ and $\beeta_0$ would be likewise be smaller than the ``naive'' errors obtained directly from the GEE model, and thus the latter may in some cases provide superior coverage. In general, we advise using the bootstrap to obtain valid standard errors for $\bbeta_0$.

\section{Simulations}

We conduct simulations to examine bias, variance, as well as coverage for the confidence intervals produced by both the GEE model-based ``naive'' standard errors (NSE)---which do not take into account uncertainty in the estimation of $\bgamma_0$ or $\beeta_0$---and compare these with bootstrap standard errors (BSE) for our model which multiplies inverse-intensity and inverse-probability of dropout weights. We hypothesize that the bias and variance when estimating $\bbeta_0$ will decrease as sample size increases.  We hypothesize that coverage will improve and that the NSEs will better match the empirical standard errors (ESE) as sample size increases. We believe that the NSEs will have nominal coverage in large samples and that they will be comparable to the BSEs even in relatively small samples---however, we would expect the bootstrap to perform better when informative dropout is more abundant or more extreme. We will refer to  the effect size of the variable affecting the dropout rate as ``informativeness,'' and we hypothesize that the bias and coverage will deteriorate in smaller samples as the dropout proportion and informativeness increase. In some cases, our weights may be highly variable due to some individuals having low visit intensity or low probability of dropout at particular times. Trimming the weights can reduce the variability at the expense of bias. As we trim the weights at lower percentiles, we expect to introduce more bias while reducing variability, in which case the NSEs and BSEs may better match the ESEs, but the estimates would be biased.

We generate visit times from Poisson process $N^*(\cdot)$ with intensity rate $\lambda_0 \exp\{\gamma_0^\top \log(\max(1+Y_i(t), 0.01))\}$ and for Scenario 1, the outcome is generated from the random-intercept model $$Y_i(t)=\beta_0+\beta_1\log(1+t) + b_i + \varepsilon_i(t),\quad b_i \sim N(0,\sigma_\phi^2),\quad \varepsilon_i(t) \sim N(0, \sigma_\varepsilon^2).$$ Scenario 1 matches the model we will use for our data example and Scenario 2 is a more complicated variant involving the reciprocal of a quadratic term (see Appendix 6.4). In both scenarios, each visit is assigned a probability of the subject experiencing informative dropout at that time based on $\text{logit}(\pi_i(t))=\eta_0+\eta_1Y_i(t)$. Informative dropout indicators are then generated for each visit time according to Bernoulli($\pi_i(t)$) and the first visit for which the indicator equaled 1 is taken to be the informative dropout time. Informativeness is controlled by varying $\eta_1$, and $\eta_0$ is adjusted accordingly to achieve various dropout proportions. Noninformative censoring is generated according to Uniform($c\tau$), where $\tau$ is a predetermined end-of-study time and $c$ is a constant which controls the proportion of noninformative censoring. Since our parameter of interest $\bbeta_0$ is a vector (two-dimensional for Scenario 1 and three-dimensional for Scenario 2), we will consider the area under the curve (AUC) as our estimand in order to simplify the simulation output. The AUC is the integral from 0 to $\tau$ of both mean models previously described; in both cases, a closed form can be computed. 

We compare our method (IIW$\times$IPW) with two alternatives. The weights for our method are computed as in (\ref{weights}), using a Andersen-Gill\cite{andersen1982cox} model for the inverse-intensities and a logistic regression model for the inverse-probabilities. Subjects either experience informative dropout, are noninformatively censored, or are uncensored in the sense that they are present up to end-of-study time $\tau$. In one of the alternatives---denoted by ``IIW'' and indicated with green in subsequent figures---weights are computed using only the inverse-intensity term and the dropout is not modelled. For the other alternative---denoted by ``IIW-NID'' and indicated with red in subsequent figures---also neglects to model dropout but further assumes we have no information about the dropout and all subjects were present up to end-of-study time $\tau$ unless they were noninformatively censored.

Bias is computed by subtracting the true AUC value from the average AUC value across all simulations. Coverage probabilities (CP) of the NSEs are computed by building 95\% confidence intervals for the AUC, then tracking the proportion which contains the true AUC across all simulations. BSEs are computed similarly, except using the 2.5th and 97.5th percentiles from a bootstrap sampling distribution instead of the model-based SEs. ESEs are computed by taking the standard deviation of the AUC across simulations. Monte Carlo SEs for the bias and coverage are computed according to Table 6 in Morris et al. \cite{morris2019using} 

For Scenario 1, we set $\gamma_0=-0.336, \bbeta_0=(16.4,-3.1)^\top, \tau = 16, c = 2$. In both scenarios, we set $\lambda_0=1, \sigma_\phi = 1, \sigma_\varepsilon = 2$. To determine the impact of informativeness, we run 48 separate simulations for each scenario---with 20,000 iterations each---for different configurations of $\eta_0$ and $\eta_1$, while also considering sample sizes of 200, 500, 1000, and 2000. We consider three values for $\eta_1$ in increasing informativeness: 0.5, 1, and 1.5. We adjust $\eta_0$ accordingly to achieve four informative dropout proportions: 20\%, 40\%, 60\%, and 80\%. We trim the weights for all three methods at the 99.9th, 99.5th and 99th percentiles. To compare the NSEs to the BSEs, we run two more simulations---with 1,000 iterations each---for only three different configurations of $\eta_0$ and $\eta_1$. These were selected from the previous twelve at the least extreme ($\eta_1=0.5$, 20\% dropout) and most extreme ($\eta_1=1.5$, 80\% dropout). We use a sample size of 200 and 100 bootstrap iterations for both configurations.

\begin{figure}[!htbp]
 \centering
 \includegraphics[width=1\textwidth]{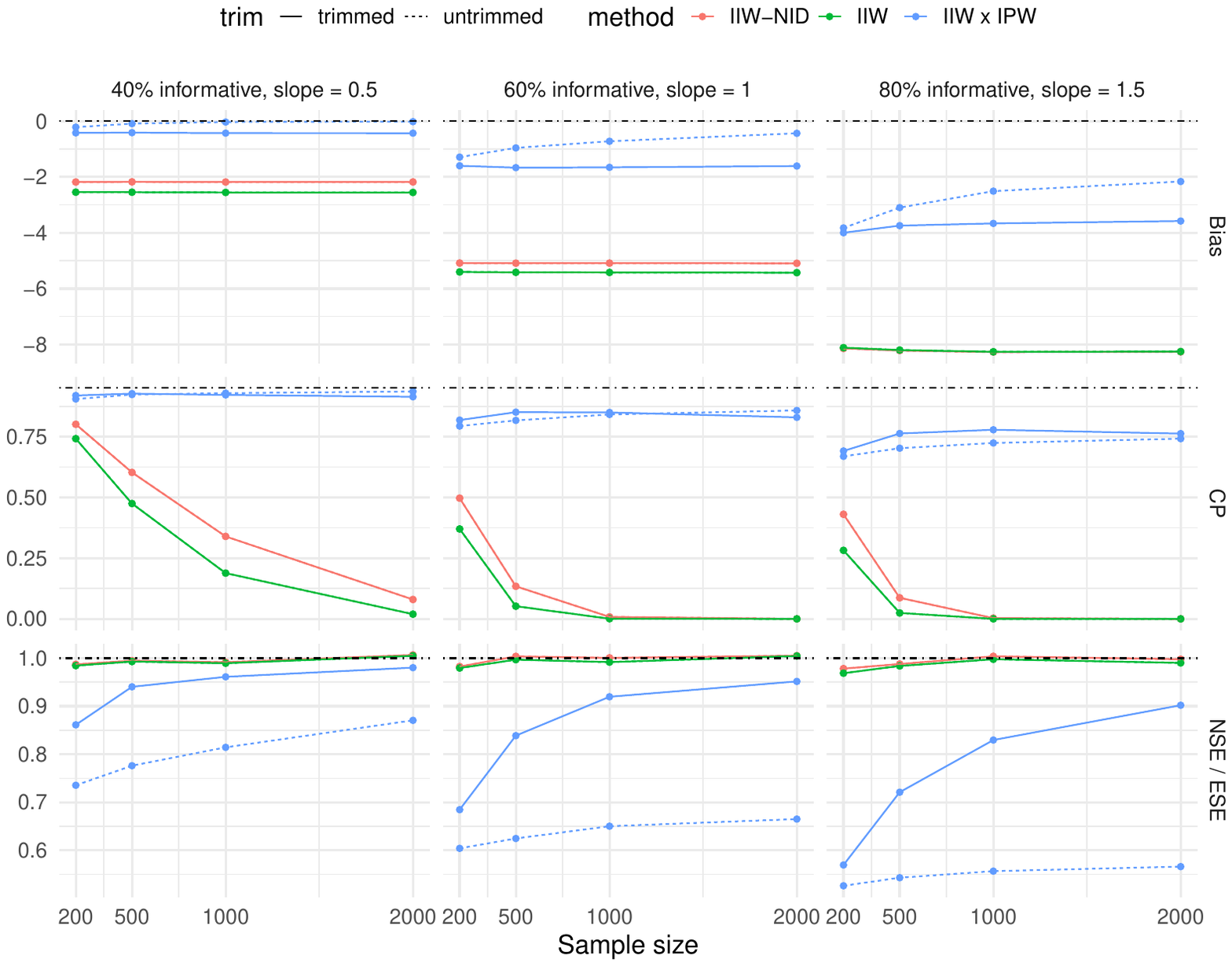}
 \caption{Bias, coverage probabilities and the ratio of naive to empirical standard error for AUC in Scenario 1 with $\eta_1 = (0.5,1,1.5)^\top$ and varying $\eta_0$ to achieve different proportions of informative dropout. All three methods are compared, with our method in blue. 99.9th percentile-trimmed weights are shown along with untrimmed weights for each case. Dashed lines represent zero bias, 95\% coverage probability, and a ratio of 1, respectively.}
 \label{stardsummary}
\end{figure}

\begin{figure}[!htbp]
 \centering
 \includegraphics[width=0.95\textwidth]{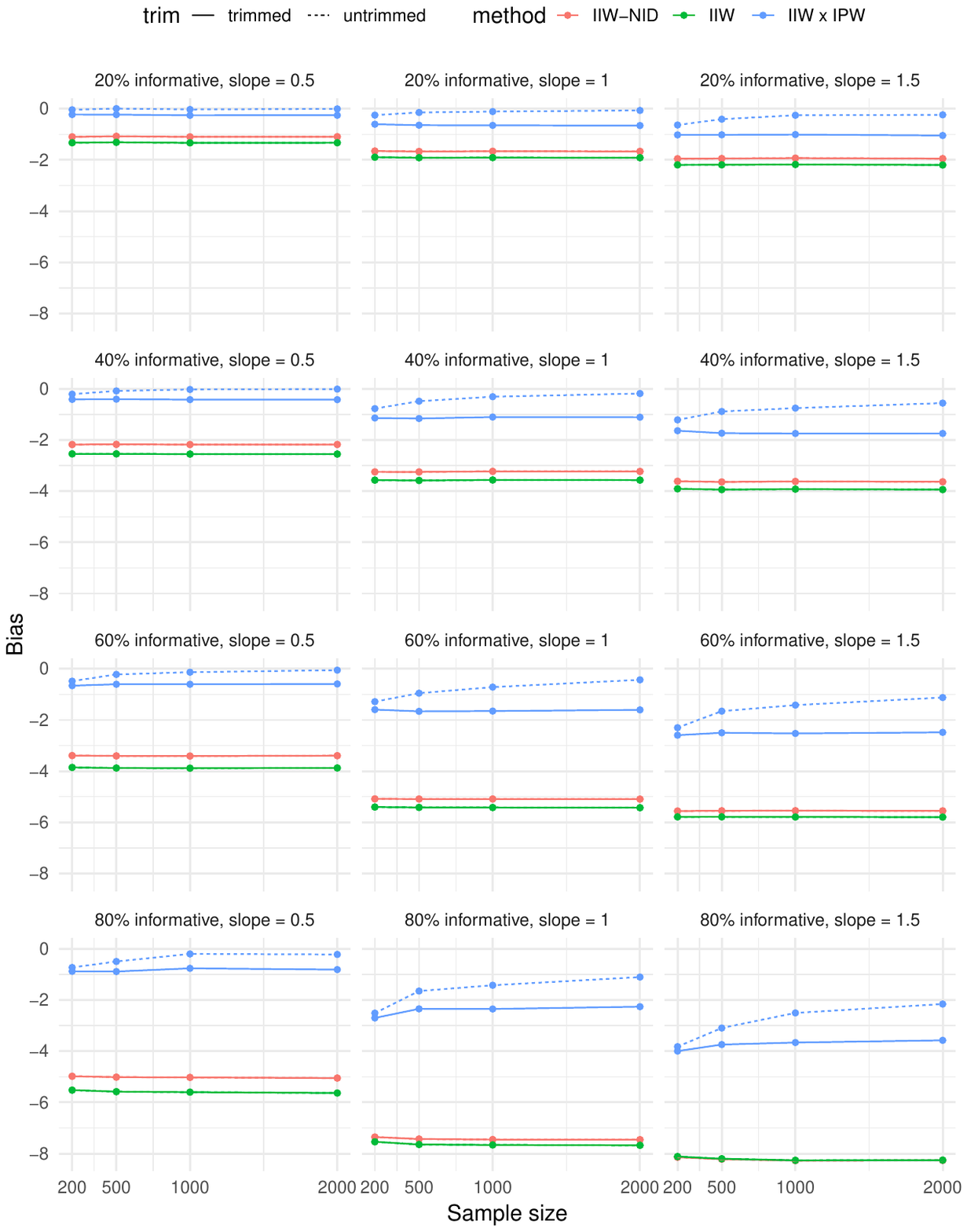}
 \caption{Bias for AUC in Scenario 1 with $\eta_1 = (0.5,1,1.5)^\top$ and varying $\eta_0$ to achieve different proportions of informative dropout. All three methods are compared, with our method in blue. 99.9th percentile-trimmed weights are shown along with untrimmed weights for each case.}
 \label{stardbias}
\end{figure}

\begin{figure}[!htbp]
 \centering
 \includegraphics[width=0.95\textwidth]{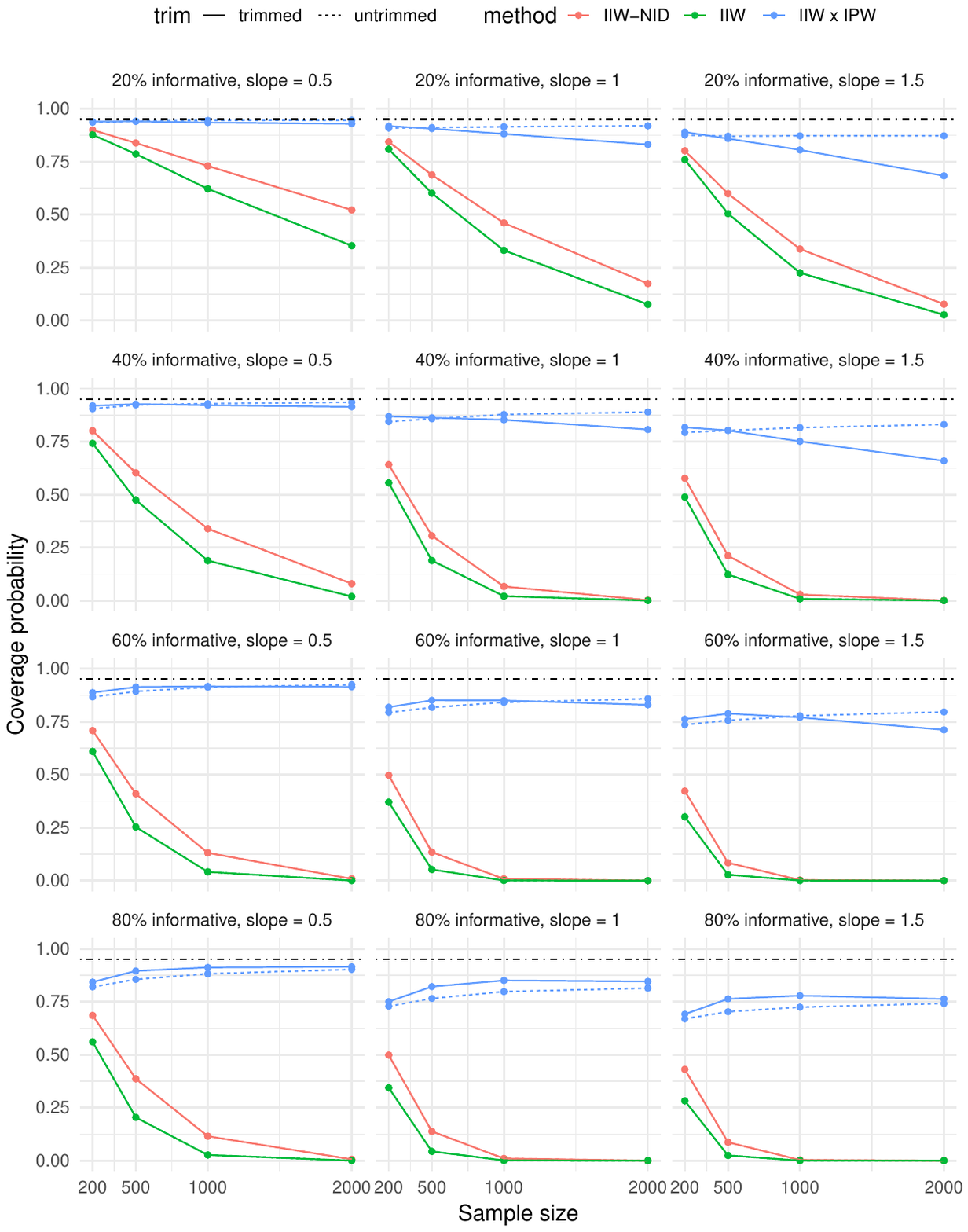}
 \caption{Coverage probabilities for AUC in Scenario 1 using naive standard errors, with $\eta_1 = (0.5,1,1.5)^\top$ and varying $\eta_0$ to achieve different proportions of informative dropout. All three methods are compared, with our method in blue. 99.9th percentile-trimmed weights are shown along with untrimmed weights. The dashed line represents 95\% coverage probability.}
 \label{stardcoverage}
\end{figure}

\begin{figure}[!htbp]
 \centering
 \includegraphics[width=0.95\textwidth]{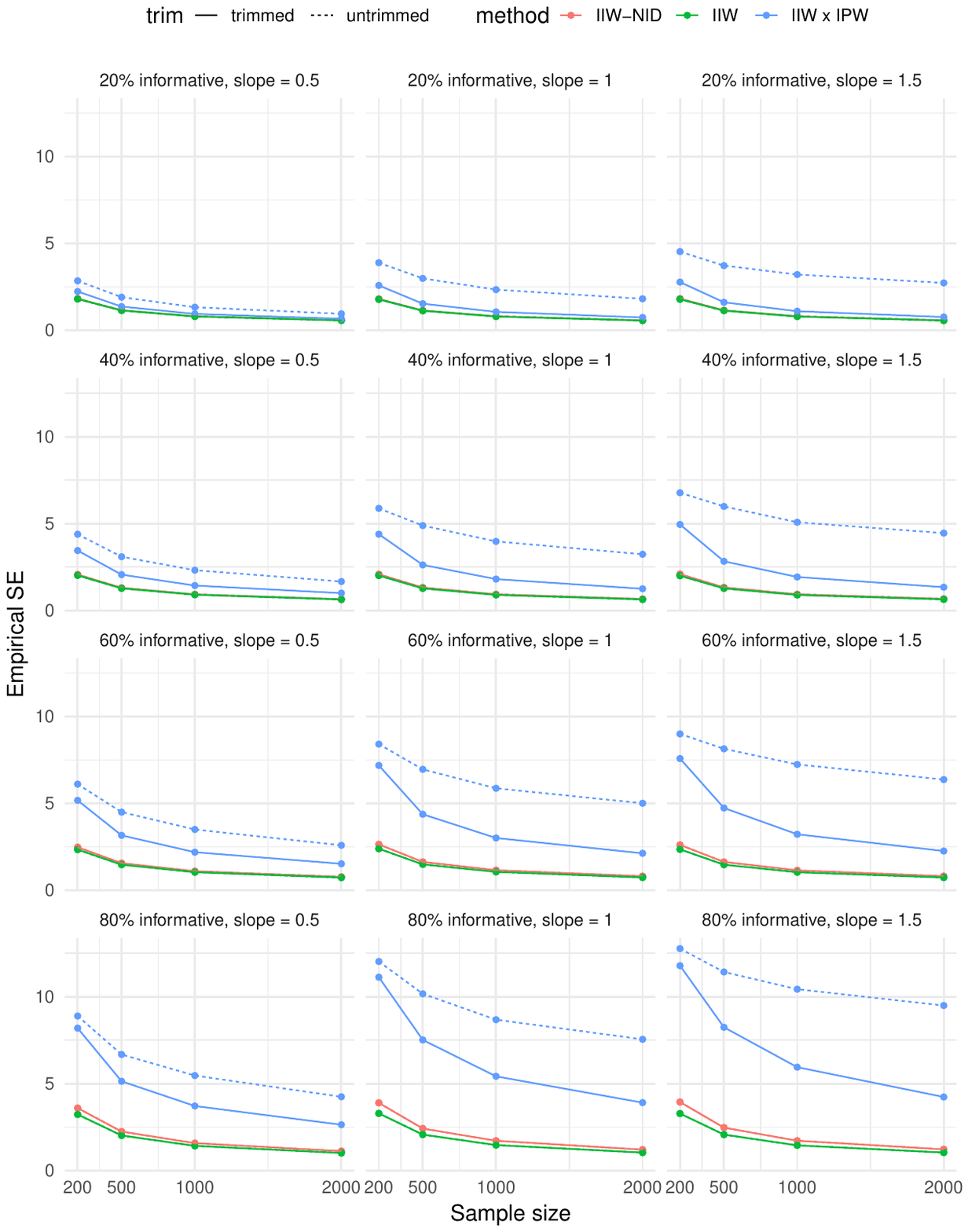}
 \caption{Empirical standard errors for AUC in Scenario 1, with $\eta_1 = (0.5,1,1.5)^\top$ and varying $\eta_0$ to achieve different proportions of informative dropout. All three methods are compared, with our method in blue. 99.9th percentile-trimmed weights are shown along with untrimmed weights.}
 \label{stardese}
\end{figure}

\begin{figure}[!htbp]
 \centering
 \includegraphics[width=0.95\textwidth]{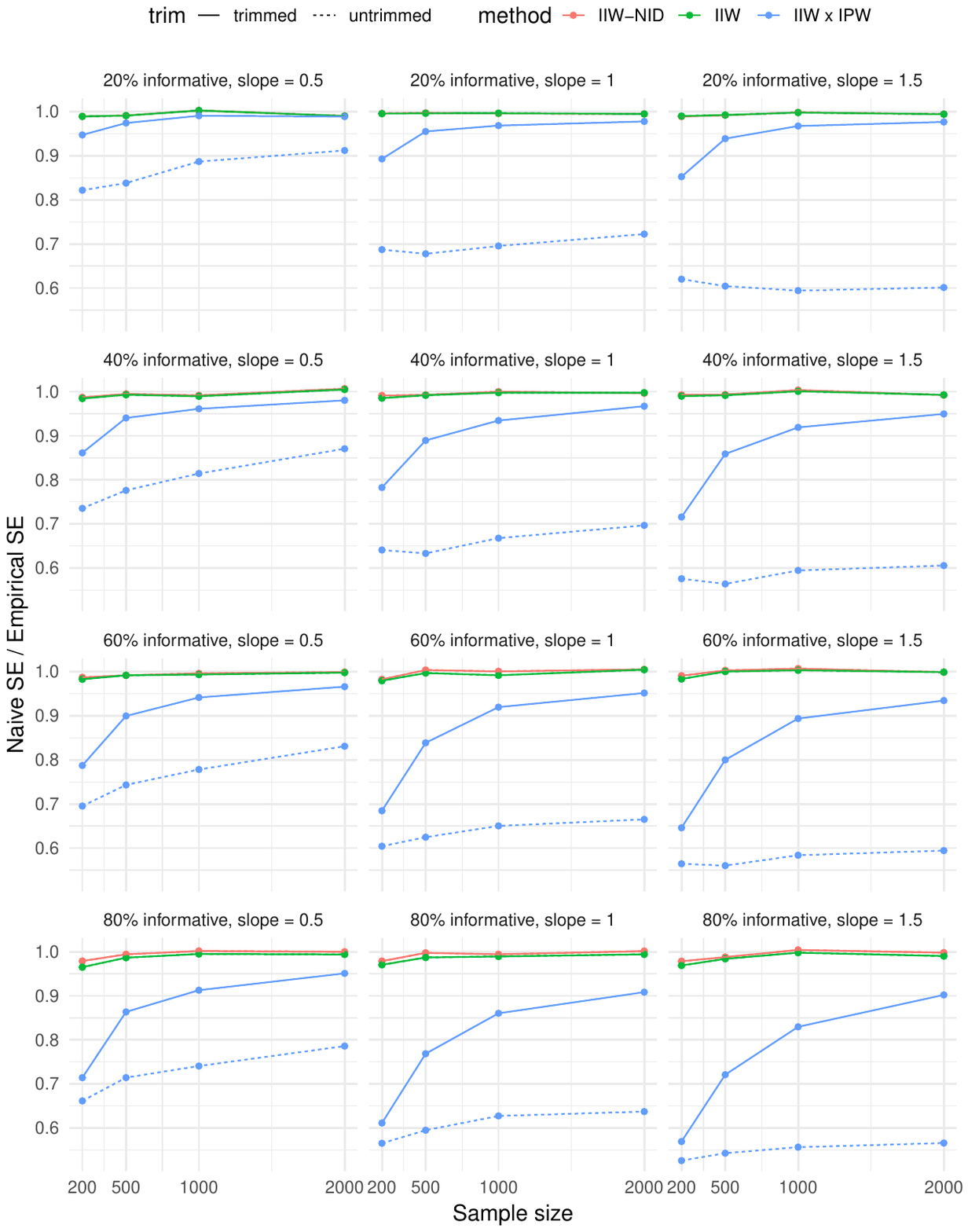}
 \caption{Ratio of naive to empirical standard errors for AUC in Scenario 1, with $\eta_1 = (0.5,1,1.5)^\top$ and varying $\eta_0$ to achieve different proportions of informative dropout. All three methods are compared, with our method in blue. 99.9th percentile-trimmed weights are shown along with untrimmed weights.}
 \label{stardratio}
\end{figure}

\begin{figure}[!htbp]
 \centering
 \includegraphics[width=0.95\textwidth]{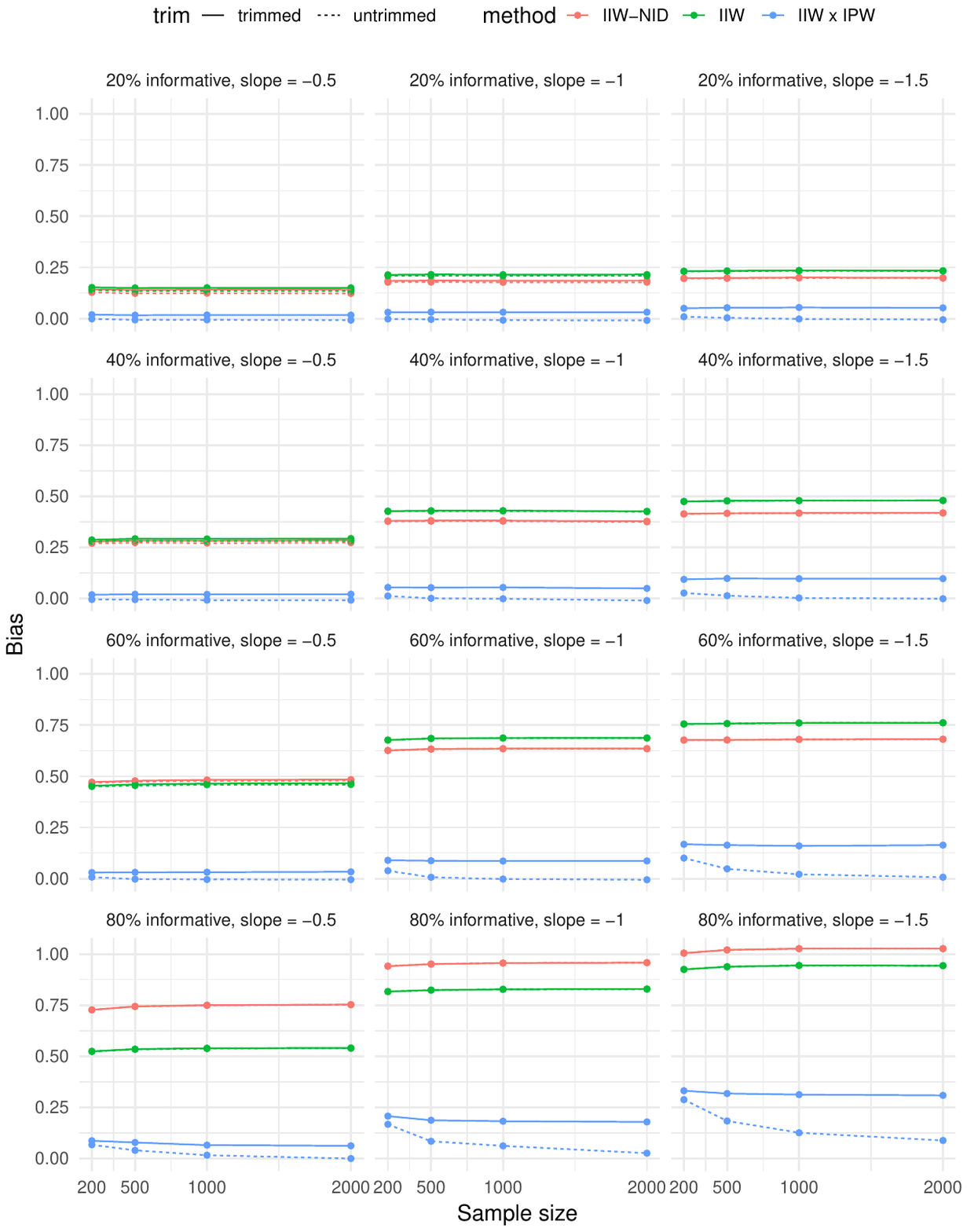}
 \caption{Bias for AUC in Scenario 2 with $\eta_1 = (-0.5,-1,-1.5)^\top$ and varying $\eta_0$ to achieve different proportions of informative dropout. All three methods are compared, with our method in blue. 99.9th percentile-trimmed weights are shown along with untrimmed weights.}
 \label{nonstardbias}
\end{figure}

\begin{figure}[!htbp]
 \centering
 \includegraphics[width=0.95\textwidth]{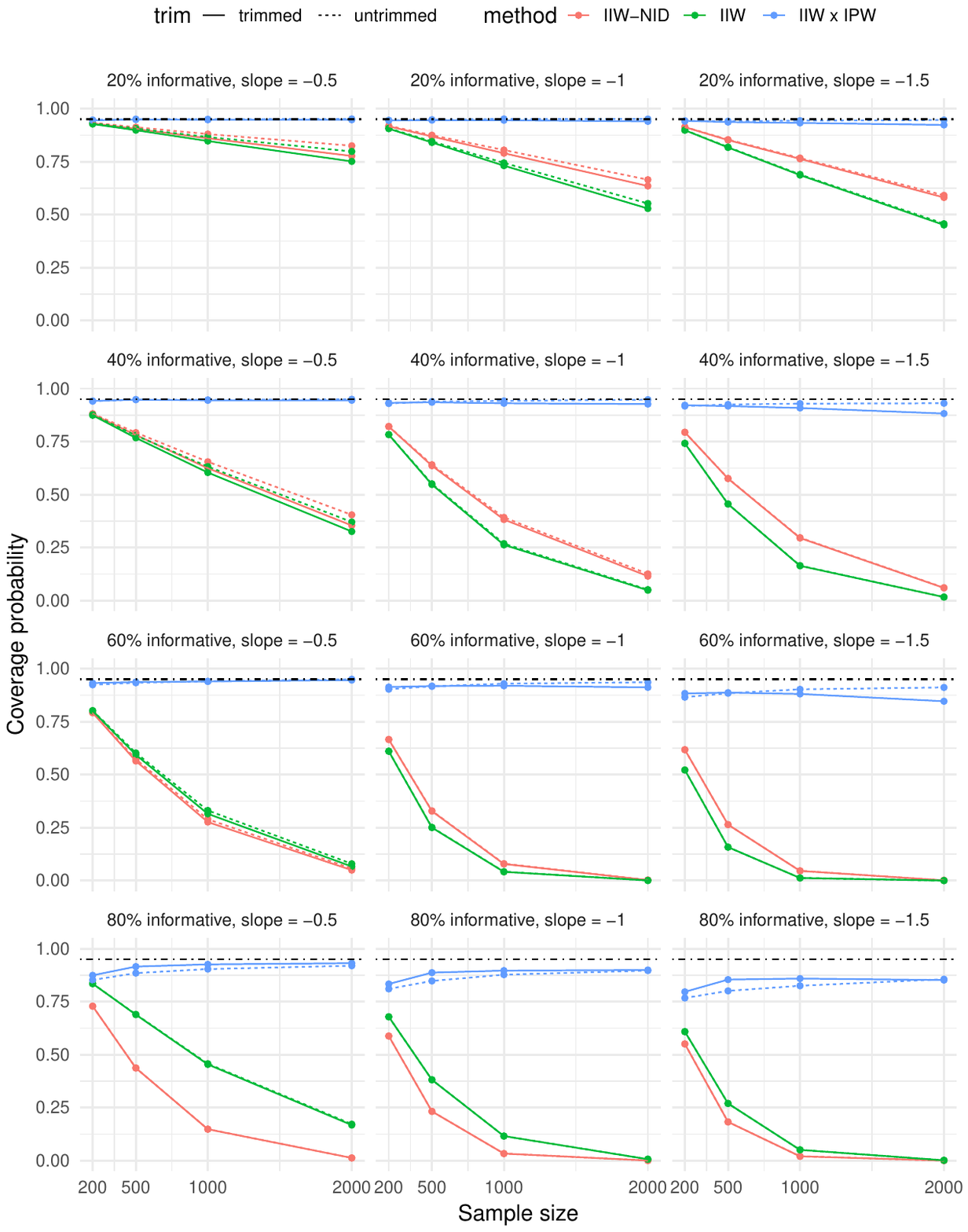}
 \caption{Coverage probabilities for AUC in Scenario 2 using naive standard errors, with $\eta_1 = (-0.5,-1,-1.5)^\top$ and varying $\eta_0$ to achieve different proportions of informative dropout. All three methods are compared, with our method in blue. 99.9th percentile-trimmed weights are shown along with untrimmed weights. The dashed line represents 95\% coverage probability.}
 \label{nonstardcoverage}
\end{figure}

\begin{figure}[!htbp]
 \centering
 \includegraphics[width=0.95\textwidth]{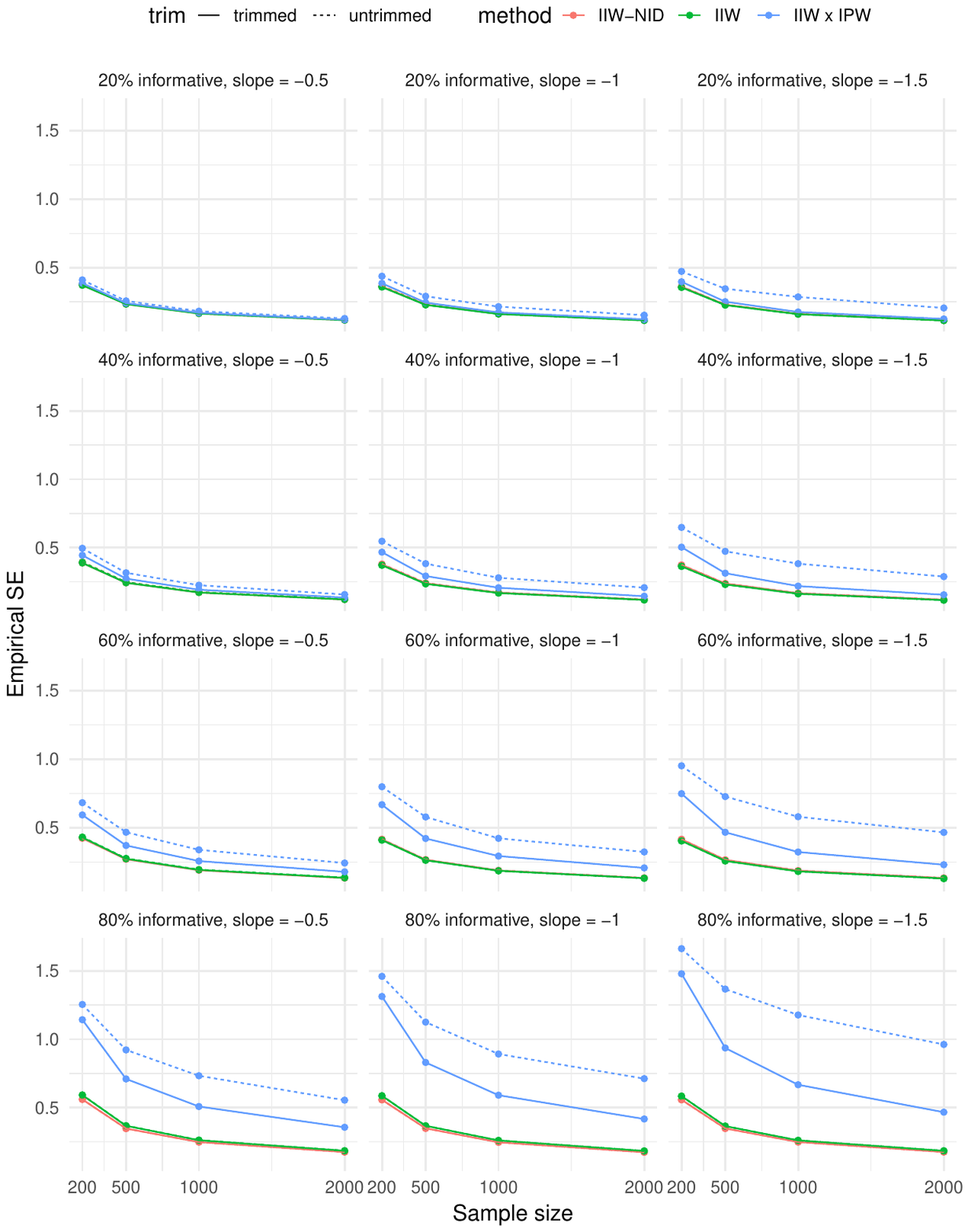}
 \caption{Empirical standard errors for AUC in Scenario 2, with $\eta_1 = (-0.5,-1,-1.5)^\top$ and varying $\eta_0$ to achieve different proportions of informative dropout. All three methods are compared, with our method in blue. 99.9th percentile-trimmed weights are shown along with untrimmed weights.}
 \label{nonstardese}
\end{figure}

\begin{figure}[!htbp]
 \centering
 \includegraphics[width=0.95\textwidth]{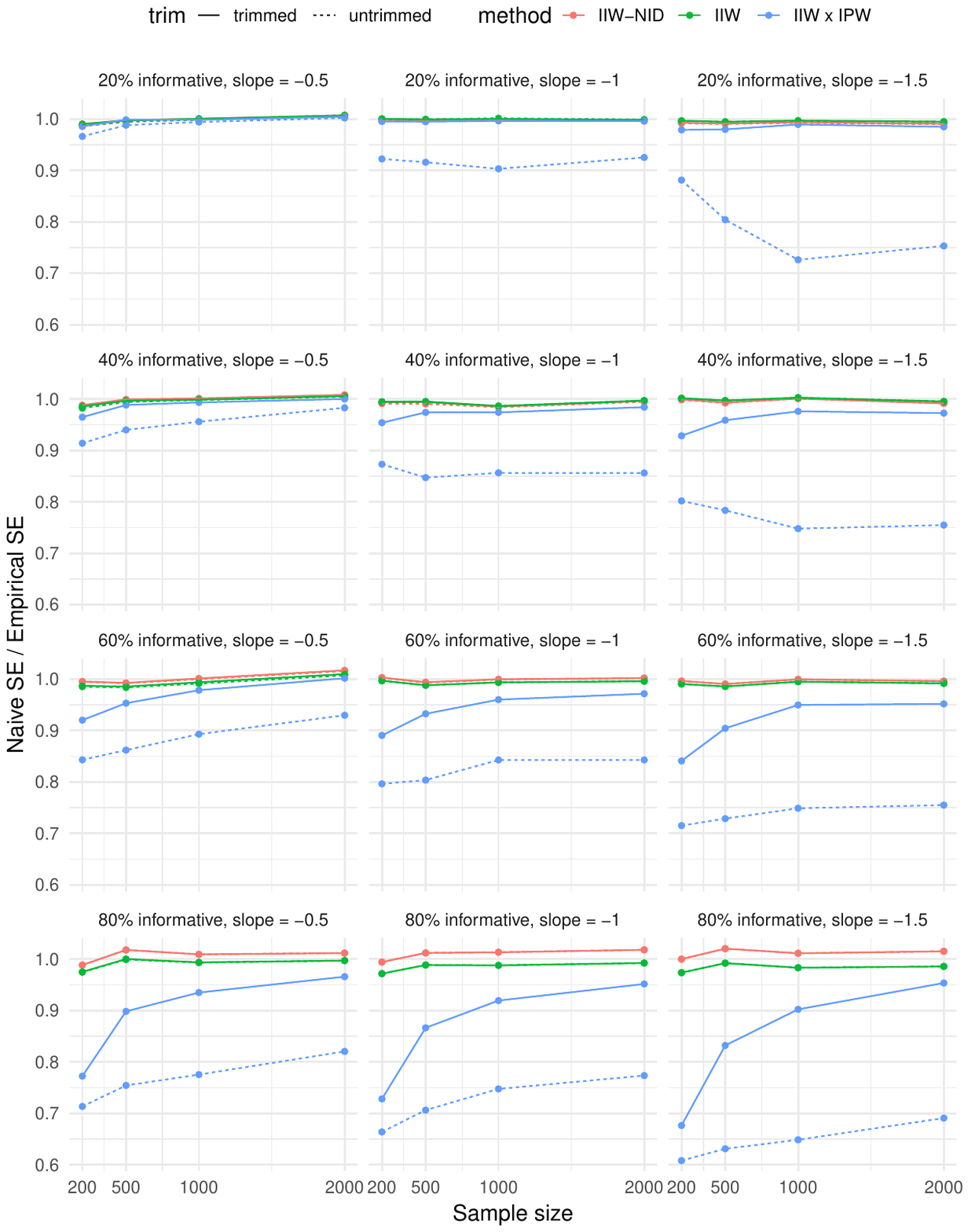}
 \caption{Ratio of naive to empirical standard errors for AUC in Scenario 2, with $\eta_1 = (-0.5,-1,-1.5)^\top$ and varying $\eta_0$ to achieve different proportions of informative dropout. All three methods are compared, with our method in blue. 99.9th percentile-trimmed weights are shown along with untrimmed weights.}
 \label{nonstardratio}
\end{figure}

We observe that the bias tends to zero for the IIW$\times$IPW method (Figures \ref{stardsummary}, \ref{stardbias} and \ref{nonstardbias}) and the ESEs become smaller as sample size increases (Figures \ref{stardese} and \ref{nonstardese}); meanwhile, the alternative methods exhibit bias even in large samples. As expected, trimming introduces bias for all configurations. Increasing dropout proportion and informativeness (while holding the other constant) leads to gradually increasing bias. Coverage improves as sample size increases for the IIW$\times$IPW method and worsens for the other two methods due to persistent bias (Figures \ref{stardsummary}, \ref{stardcoverage} and \ref{nonstardcoverage}). While coverage is adequate in the low and medium settings, achieving 95\% coverage becomes more difficult with increasing informativeness, necessitating larger sample sizes. Trimming tends to lead to worse coverage due to bias, although it performs as well or better in more extreme parameter cases. The ratio of NSEs to ESEs tends closer to 1 with larger sample size for the IIW$\times$IPW method, except in a few cases of high informativeness; this is due to the presence of extreme weights (Figures \ref{stardsummary}, \ref{stardratio} and \ref{nonstardratio}). In such cases, trimming leads to the NSEs better representing the underlying variability and appears to provide more of a benefit as dropout proportion and informativeness increase. NSEs and ESEs are similar for the other methods at all sample sizes.

When considering a relatively small sample size of 200, we observe for both simulation scenarios that in the least extreme case the coverage is adequate, but worsens as the extent of the informative dropout and the level of informativeness increase, due in part to underestimation of the standard errors. This can be in part mitigated by using the bootstrap, which better recovers the empirical standard errors, especially when used in conjunction with trimming in the high dropout and high informativeness scenario, at the expense of increased bias (Tables \ref{bootsummary}, \ref{scenario1} and \ref{scenario2}). We therefore recommend the bootstrap standard errors and potentially trimming the weights---if the estimated weights are very extreme---for situations with high informative dropout rates, or when the outcome has a large impact on the dropout.

\begin{table}[!htbp]
\centering
\begin{tabular}{lllrrrrrr}
  \toprule
 \% inf & Slope & Trim & Bias & Emp SE & Naive SE & Naive CP & Boot SE & Boot CP \\
 \midrule
 20\% & 0.5 & None & -0.19 & 2.8 & 2.3 & 0.94 & 2.5 & 0.96 \\
  &  & 99.9\% & -0.28 & 2.1 & 2.1 & 0.95 & 2.4 & 0.97 \\
  &  & 99.5\% & -0.49 & 1.9 & 1.9 & 0.94 & 2.0 & 0.95 \\
  &  & 99\% & -0.60 & 1.8 & 1.9 & 0.95 & 1.9 & 0.95 \\
 \midrule
 60\% & 1.0 & None & -1.3 & 8.4 & 5.0 & 0.78 & 5.6 & 0.84 \\
  & & 99.9\% & -1.6 & 7.1 & 4.9 & 0.81 & 5.4 & 0.87 \\
  &  & 99.5\% & -2.6 & 4.4 & 3.9 & 0.83 & 4.6 & 0.88 \\
  &  & 99\% & -3.0 & 3.6 & 3.4 & 0.81 & 3.7 & 0.84 \\
 \midrule
 80\% & 1.5 & None & -3.4 & 14 & 6.8 & 0.69 & 8.2 & 0.78 \\
  &  & 99.9\% & -3.6 & 12 & 6.8 & 0.70 & 8.1 & 0.80 \\
  &  & 99.5\% & -4.7 & 7.2 & 5.9 & 0.75 & 7.3 & 0.83 \\
  &  & 99\% & -5.4 & 5.7 & 5.1 & 0.73 & 5.9 & 0.78 \\
 \bottomrule 
\end{tabular}
\caption{AUC for Scenario 1 with $n_\text{sim}=1000$ and $n=200$ when bootstrapping standard errors for the IIW$\times$IPW method. \% inf is the percentage of subjects who were informatively censored and the slope $\eta_1$ controls ``informativeness.'' Bias is presented along with empirical standard errors, and we further present naive and bootstrap standard errors along with their corresponding coverage probabilities (CP). Four levels of trimming are shown.}
\label{bootsummary}
\end{table}


\section{Data Example}

We will demonstrate the impact of ignoring informative dropout through the Sequenced Treatment Alternatives to Relieve Depression (STAR*D) trial.\cite{rush2006acute, trivedi2006evaluation} The objective of the study was to determine the efficacy of various treatments throughout multiple stages of randomization, referred to in the study as ``levels''---if a patient did not respond well to a treatment, they would move on to a new level and undergo a new regimen. For our purposes, we will focus on Level 1, during which subjects are prescribed Citalopram; the protocol recommended visits at 2, 4, 6, 9, and 12 weeks, with an optional 14-week visit if required \cite{trivedi2006evaluation}. The outcome of interest is the Quick Inventory of Depressive Symptomatology (QIDS) score (clinician-rated), which serves as a measure for the severity of major depressive disorder (MDD)---it ranges from 0 to 27, with higher scores indicating more severe depression. Remission was defined as the first instance of a QIDS score less than or equal to 5. \cite{rush2006acute} 

Our aim is to investigate the longitudinal trajectory of the QIDS score. Many subjects deviated from the protocol, visiting earlier or later than prescribed; an increased rate of visits may indicate inadequate treatment response. All patients who did not achieve remission by the end of Level 1 were advised to enter the next level; however, patients were also advised do so before the 12-week mark if they experienced intolerable side effects or if they had not responded to treatment after approximately nine weeks.\cite{trivedi2006evaluation} Thus, the structure of this trial could lead to both informative visits and informative dropout. 

Several demographic variables had missing data, so we first used multiple imputation by chained equations (MICE)\cite{van2011mice} to obtain 20 imputed datasets. To specify the intensity model as correctly as possible, we modelled the visit process as days rather than weeks. We determined that a patient's next recommended visit should be in 2 weeks if they have just had a visit or if they are at most 36 days into the trial, and in 3 weeks thereafter. We count the number of days up to every visit, and define a variable by taking the difference between the current day and the recommended visit time for every subject; we cap this difference at 7 days and discretize the variable as: not yet due for a visit; within 3 days of recommended visit time; and more than 3 days late. We then determine the number of visits each patient is expected to have had by a certain time: no visits before day 14; 1 visit before day 35; 2 visits before day 49; 3 visits before day 70; 4 visits before day 91; and 5 visits thereafter. This is designed to match the protocol, leaving a few days to spare to account for weekends. We define a second variable by taking the difference between each subject's actual number of visits by each time and their expected number of visits; we discretize this variable as: 2 or more fewer visits than there should have been; 1 fewer visit than there should have been; right number of visits; 1 or more visits than specified in the protocol.

\begin{figure}[!htbp]
 \centering
 \includegraphics[width=1\textwidth]{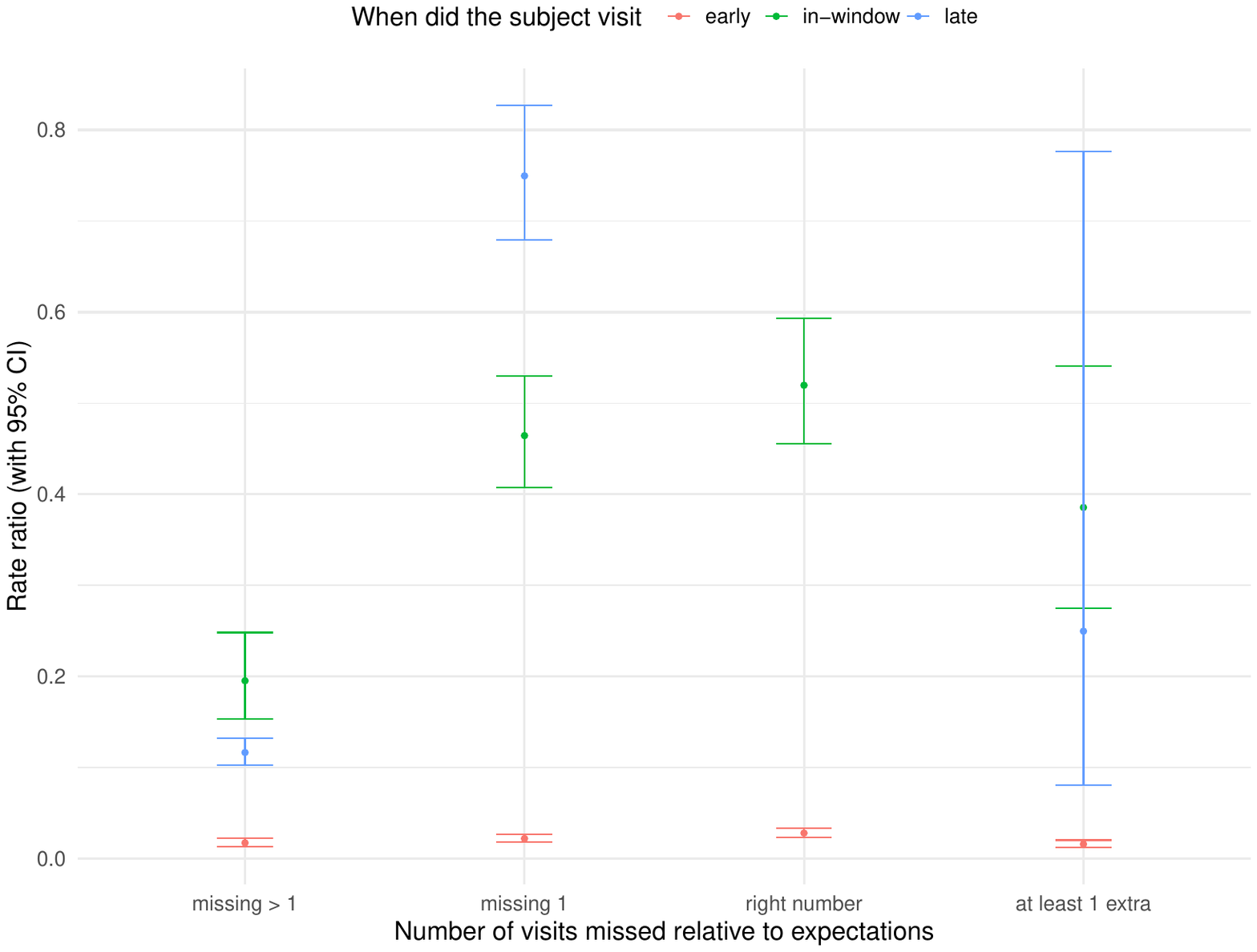}
 \caption{Intensity rate ratios (along with 95\% confidence intervals) associated with the interaction between the discretized variables representing when the subject visited relative to recommendation and the difference between actual and expected number of visits, respectively.}
 \label{visitcoefs}
\end{figure}

\begin{figure}[!htbp]
 \centering
 \includegraphics[width=1\textwidth]{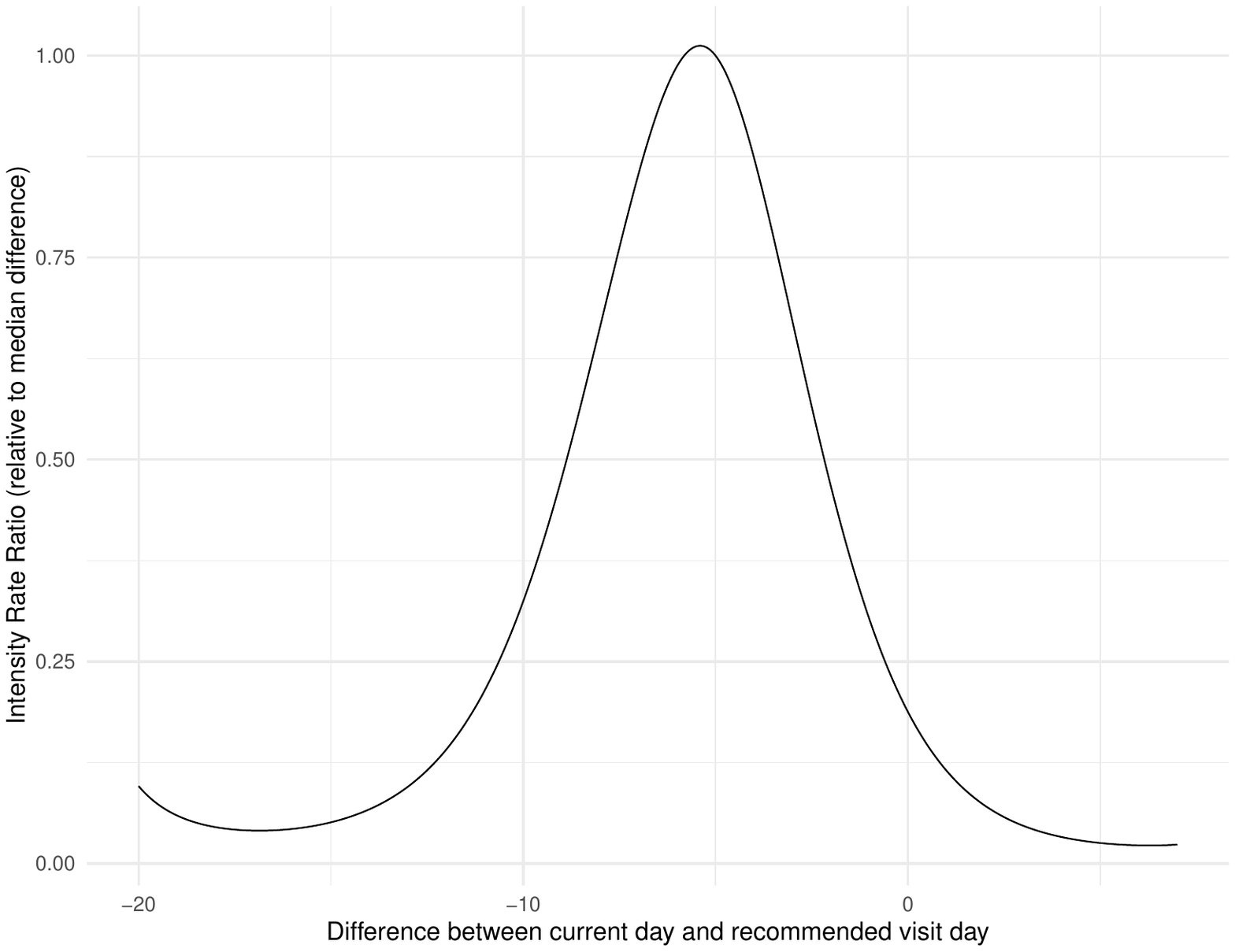}
 \caption{Predicted intensity rate ratios across the range of differences between current day and recommended visit day (relative to median difference), based on the estimated B-spline coefficients.}
 \label{splineplot}
\end{figure}

We fit a Poisson GLM with the visit indicator as the outcome; this allows the visit intensity to change on a daily basis as a function of the continuous difference between each day and the recommended time, as we suspect that subjects are increasingly likely to visit as the scheduled day nears. To accommodate potential nonlinear trends, we use a set of four cubic B-spline basis functions; the estimated trajectory is shown in Figure \ref{splineplot}. We include the interaction between the two discretized variables to account both for whether subjects are nearing a recommended time and how much they have deviated from the schedule in the past (Figure \ref{visitcoefs}). In order to capture the effect of visiting early and having had fewer visits than expected up to that point, we include an interaction term of whether the latter variable is larger than 1 and the former is less than 0. We then include an interaction between the QIDS score at the previous visit and an indicator of whether or not they are at most 10 days into their next visit sequence, as well as a variable indicating whether they are 7 days into their next visit sequence. We also include various demographic variables in the model (Table \ref{democoefstable}), described in detail in Table \ref{demotable}.

\begin{figure}[!htbp]
 \centering
 \includegraphics[width=1\textwidth]{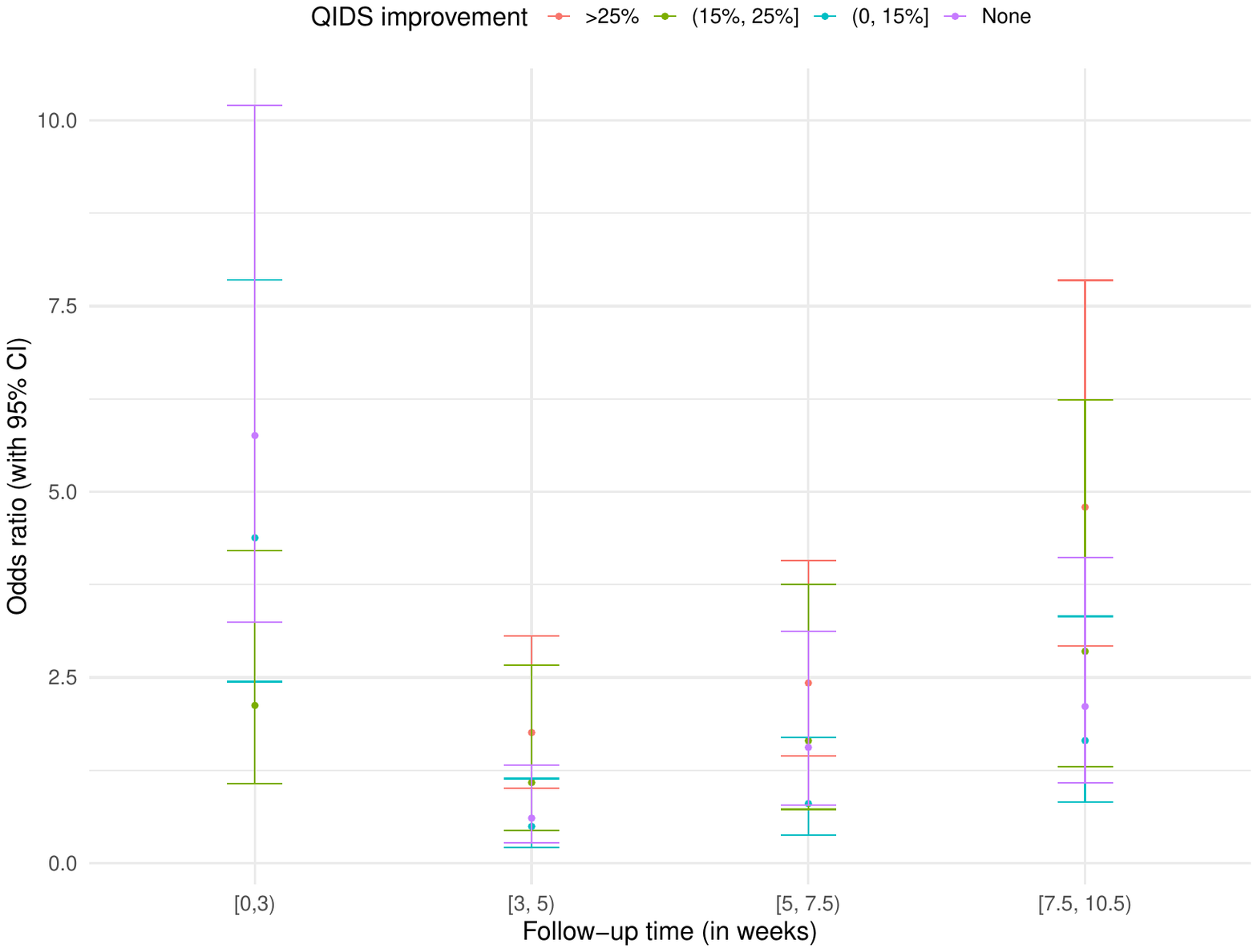}
 \caption{Odds ratio estimates (with 95\% confidence intervals) from the dropout model for every combination of discretized follow-up time and percent improvement in QIDS score relative to baseline.}
 \label{dropcoefs}
\end{figure}

We define our informative dropout indicator by using a database which tracked the time each patient left Level 1 before week 12; their reasons for doing so were also recorded in the data. If they dropped out for a reason that could potentially have been related to the outcome---as previously outlined---we record them as an informative dropout; otherwise, we deem them to have been uncensored. We fit a logistic regression model to the weekly longitudinal data with the indicator as the outcome; for the predictors, we use the baseline QIDS score for each patient, as well as the interaction between the discretized percentage change in QIDS score from baseline at any given time and the discretized time variable (see Table \ref{dropcoefs}). We then compute two sets of weights: one only with inverse-intensities (IIW) and another based on (\ref{weights}), which we label IIW$\times$IPW. In both cases, we stabilize the weights with a locally constant loess of the inverse-weights over time, and trim both sets of weights at the 1st and 99th percentiles to reduce variability. We then fit a weighted GEE and model time as a continuous variable with a set of four cubic B-spline basis functions. We use bootstrap standard errors to construct 95\% confidence intervals and use Rubin's rules to pool the predictions across the imputed datasets.

When comparing the GEE models with the IIW and IIW$\times$IPW weights, we observe that the estimated trajectory of the mean QIDS score is similar throughout the first 4 weeks for both the IIW-GEE and the IIW$\times$IPW-GEE. However, our method estimates a higher mean QIDS score in later follow-up periods, though the methods begin to converge near the 12-week mark. This implies that when failing to account for informative dropout, we may slightly overestimate the improvement in the QIDS score trajectory; this seems reasonable, since this would mean ignoring that some patients may have ended treatment early as their condition did not sufficiently improve.

\begin{figure}[!htbp]
 \centering
 \includegraphics[width=1\textwidth]{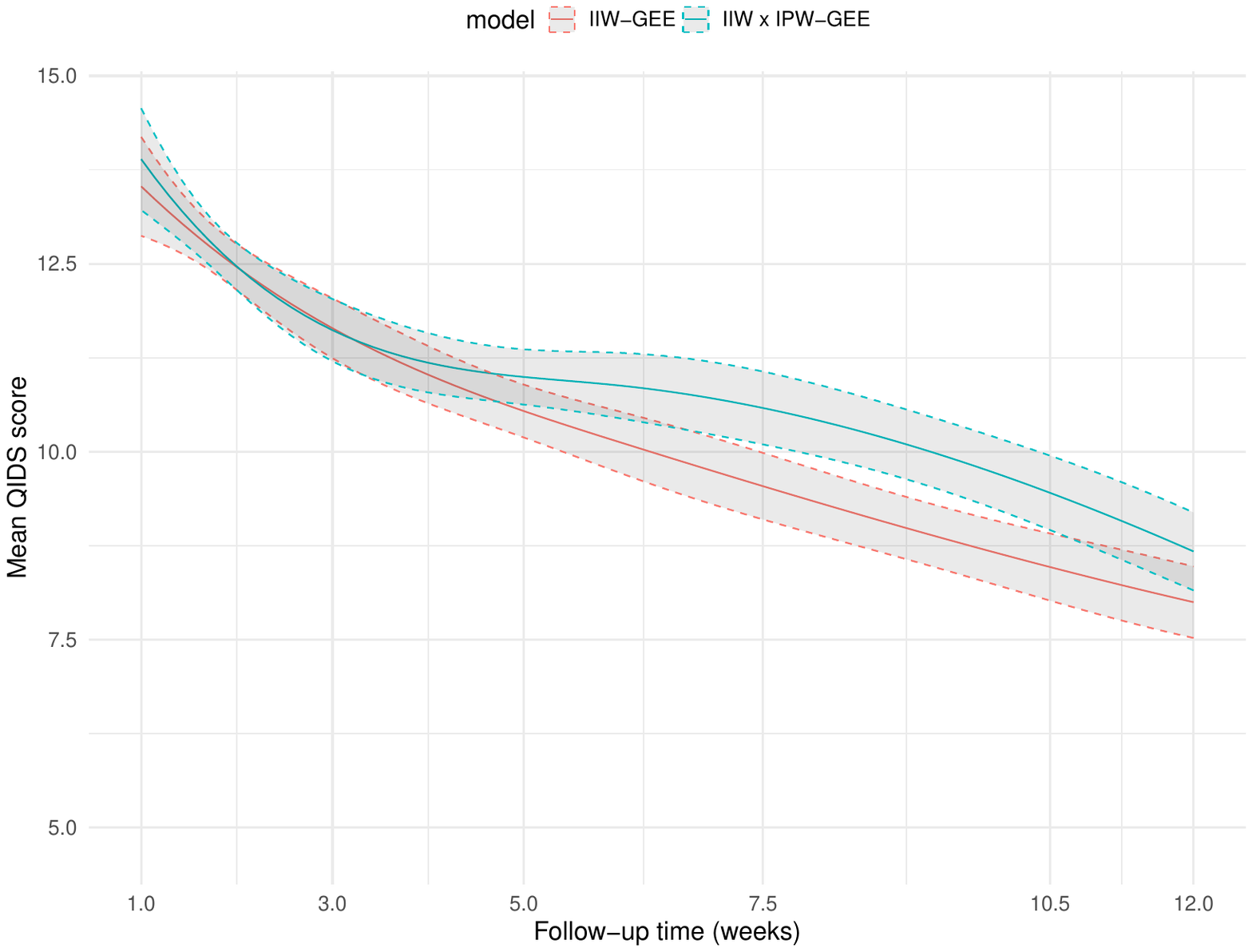}
 \caption{Estimated longitudinal trajectories for the mean QIDS score with 95\% bootstrap confidence intervals. We compare two GEE models, one only with inverse-intensity weights---ignoring potentially informative dropout---and the other accounting for informative dropout by further multiplying by the inverse-probability of not having dropped out.}
 \label{QIDS}
\end{figure}

\section{Discussion}

We find that when informative dropout is unaddressed---either by assuming all subjects are under observation until the end of the study or by acknowledging the dropout time, but only using inverse-intensity weights---this leads to biased estimates for the longitudinal model parameters. Our method, which multiplies the inverse-intensity and inverse-probability weights, reduces this bias. In the STAR*D analysis, we find that accounting for the dropout results in a higher mean QIDS score trajectory. We generally recommend using bootstrap standard errors, especially in situations of abundant informative dropout or informativeness.

In the STAR*D study, informative dropout could only occur at a visit time; patients could only be advised to enter the next level during one of their visits. We had to thus specify our weights more carefully in order to avoid positivity violations. However, computing the integral in (\ref{dropoutexplicit}) may be difficult in practice. In the simulations, we simplified this procedure by assuming a constant hazard over time; in the STAR*D dataset this was not realistic. We dealt with this by using a flexible piecewise hazard model in which the hazard was allowed to smoothly change on a daily basis. This had the added benefit of enabling us to simply add the rates instead of having to integrate. 

Throughout this paper, we assumed that the informative dropout is observable; for example, in the STAR*D data, the dropout was informative by design. Our method is unable to deal with latent dropout, as we require the explicit dropout times in order to compute the weights. Latent dropout could potentially be handled with artificial censoring, where we would censor subjects after they have missed two visits, or gone a prespecified period of time without a visit. For example, when patients are treated for lupus, they should be seen every six months; in this case, we could informatively censor them if they have gone two years without a visit. Both the visit intensity and dropout models must be correctly specified; in order to satisfy the conditional independence assumptions, care must be taken to include the variables associated with both visiting and dropping out, respectively, as well as with the outcome. However, if this can be achieved, our method is flexible in that it allows for the inclusion of any combinations of variables in the models, which is particularly useful when the outcome has an impact on visiting or dropout.

Our investigations confirm assertions by Coulombe et al.\cite{coulombe2024multiply} and Tompkins et al.\cite{tompkins2024flexible} that neglecting to adjust for informative dropout may lead to bias. Coulombe et al. suggested that multiplying by the censoring weights could be a potential solution; we show with mathematical derivations how this can work for our setting and confirmed the results via simulations. While Tompkins et al. provided simulations results in which they aimed to determine the impact of multiplying by censoring weights, their results seemed to suggest that these weights were not particularly helpful, which we suspect may be due to their censoring process not having any associated time-dependent covariates. Our simulations demonstrate that when the dropout is outcome and time-dependent, failing to adjust for informative dropout as we propose may have a dramatic impact.

An augmented variant of our estimating equation, such as those explored by Coulombe et al., \cite{coulombe2024multiply} could be explored which may yield some flexibility in the misspecification of at least one of the three models. Dealing with latent dropout in the context of GEEs is an area worth further investigation, either through artificial censoring or by making assumptions about the missingness mechanism and modelling it jointly. We have also assumed that a potential competing event is distinct from the dropout event; it may be worth exploring scenarios in which the dropout itself may prevent the outcome from being measured. 

Our work suggests that when a longitudinal study study exhibits dropout, particular attention should be paid to whether the outcome may be associated with the the dropout. If the interest is in estimating a marginal trajectory and the dropout is clearly defined and available in the data, our method allows for a straightforward approach within the GEE framework to reduce bias in the estimates and provide better coverage.

\section*{Conflict of interest}

The authors declare no potential conflict of interests.

\section*{Data Availability}
The Sequenced Treatment Alternatives to Relieve Depression study used in the preparation of this manuscript were obtained from the National Institute of Mental Health (NIMH) Data Archive (NDA). NDA is a collaborative informatics system created by the National Institutes of Health to provide a national resource to support and accelerate research in mental health. Dataset
identifier (s): [NIMH Data Archive Collection ID (s) or NIMH Data Archive Digital Object Identifier (doi://10.15154/1527974)]. This manuscript reflects the views of the authors and may not reflect the opinions or views of the NIH or of the submitters submitting original data to NDA.

\clearpage
\bibliographystyle{ama}
\bibliography{citation}

\section{Appendix}
\subsection{STAR*D demographic variables}

\begin{table}[H]
\centering
\scalebox{0.8}{
\begin{tabular}{lll}
  \toprule
  Characteristic & Number (\%) & Missing (\%) \\ 
  \midrule
  \textbf{Sex} & & 0 (0.0\%) \\
  Female & 2530 (62.6\%) & \\
  Male & 1509 (37.4\%) &  \\
  \midrule
  \textbf{Age in decades} & & 2 (0.0\%)\\
  Median [Min, Max] & 4.05 [1.81, 7.57] & \\
  \midrule
  \textbf{Current residence} & & 4 (0.1\%) \\
  Apartment or condominium & 1408 (34.9\%) & \\
  Detached house/rowhouse or townhouse/mobile home & 2505 (62.0\%) & \\
  Other & 122 (3.0\%) & \\  
  \midrule
  \textbf{Current marital status} & & 4 (0.1\%) \\
  Married & 1332 (33.0\%) & \\
  Unmarried & 2703 (66.9\%) & \\
  \midrule
  \textbf{Total number of persons in household} & & 8 (0.2\%) \\
   $\leq 2$ & 2045 (50.6\%) & \\
   $>2$ & 1986 (49.2\%) & \\
   \midrule
   \textbf{Currently a student} & & 4 (0.1\%) \\
   No & 3442 (85.2\%) & \\
   Yes & 593 (14.7\%) & \\
   \midrule
   \textbf{Current employment status} & & 37 (0.9\%) \\
   Employed & 2291 (56.7\%) & \\
   Retired & 233 (5.8\%) & \\
   Unemployed & 1478 (36.6\%) & \\
   \midrule
   \textbf{Currently do volunteer work} & & 8 (0.2\%) \\ 
   No & 3452 (85.5\%) & \\
   Yes & 579 (14.3\%) & \\ 
   \midrule
   \textbf{On medical or psychiatric leave} & & 9 (0.2\%) \\
   No & 3718 (92.1\%) & \\
   Yes & 312 (7.7\%) & \\
   \midrule
   \textbf{Has private insurance} & & 85 (2.1\%) \\
   No & 1932 (47.8\%) & \\
   Yes & 2022 (50.1\%) & \\
   \midrule
   \textbf{Better able to enjoy things} & & 16 (0.4\%) \\
   Agree & 3690 (91.3\%) & \\
   Disagree & 35 (0.8\%) & \\
   Neutral & 298 (7.4\%) & \\
   \midrule
   \textbf{Better able to make important decisions} & & 16 (0.4\%) \\
   Agree & 3650 (90.4\%) & \\
   Disagree & 46 (1.1\%) & \\
   Neutral & 327 (8.1\%) & \\
   \midrule
   \textbf{Impact of your family and friends} & & 18 (0.4\%) \\
   Difficult & 900 (22.3\%) & \\
   Helpful & 2345 (58.1\%) & \\
   Neutral & 776 (19.2\%) & \\
   \midrule
   \textbf{Number of decades in formal education} & & 12 (0.3\%) \\
   Median [Min, Max] & 1.30 [0, 2.70] & \\   
   \bottomrule
\end{tabular}}
\caption{Other current residence category includes rooming house or hotel; retirement complex or senior nursing; healthcare facility or nursing home; and homeless. Not married includes never married; living with someone; separated; divorced; and widowed. Enjoyment was captured via responses to the statement ``If I can get the help I need from a doctor, I believe that I will be much better able to enjoy things.'' Making decisions was captured via responses to the statement ``If I can get the help I need from a doctor, I believe that I will be better able to make important decisions.''}
\label{demotable}
\end{table}

\subsection{Intensity model output}

\begin{table}[!htbp]
\centering
\scalebox{0.95}{
\begin{tabular}{lll}
  \toprule
 Predictor & Level or Units & IRR (95\% CI) \\ 
  \midrule
\textbf{Sex} & Male & 1.017 (0.982, 1.054) \\ 
\midrule
\textbf{Age} & Decades & 1.033 (1.017, 1.049) \\ 
\midrule
\textbf{Residence} & House & 0.999 (0.959, 1.039) \\ 
Reference: Apartment & Other & 0.955 (0.891, 1.023) \\
\midrule
\textbf{Marital status} & Married & 0.936 (0.899, 0.974) \\ 
\midrule
$>2$ \textbf{people in household} & & 0.931 (0.896, 0.966) \\ 
\midrule
\textbf{Student} & No & 1.030 (0.979, 1.084) \\ 
\midrule 
\textbf{Employment} & Retired & 0.957 (0.883, 1.036) \\ 
Reference: Employed & Unemployed & 0.965 (0.928, 1.004) \\ 
\midrule 
\textbf{Volunteer} & Yes & 1.022 (0.975, 1.072) \\ 
\midrule 
\textbf{Medical/psychiatric leave} & Yes & 0.935 (0.875, 0.999) \\ 
\midrule
\textbf{Has private insurance} & Yes & 1.016 (0.977, 1.056) \\ 
\midrule
\textbf{Enjoyment} & Disagree & 0.998 (0.808, 1.234) \\ 
Reference: Agree & Neutral & 1.034 (0.959, 1.115) \\ 
\midrule
\textbf{Decision-making} & Disagree & 1.010 (0.838, 1.217) \\ 
Reference: Agree & Neutral & 0.962 (0.895, 1.034) \\ 
\midrule
\textbf{Impact of family \& friends} & Helpful & 1.007 (0.965, 1.05) \\ 
Reference: Difficult & Neutral & 0.980 (0.929, 1.034) \\ 
\midrule
\textbf{Education} & Decades & 1.281 (1.209, 1.357) \\ 
\midrule 
\textbf{QIDS score and visit profile} & & \\
7 days into next visit sequence & & 2.702 (2.364, 3.088) \\ 
1 visit more than expected \& early visit & & 3.955 (2.456, 6.369) \\ 
QIDS score at previous visit \& $\geq11$ days into next visit sequence & & 1.042 (1.038, 1.045) \\ 
QIDS score at previous visit \& $<11$ days into next visit sequence & & 0.920 (0.914, 0.926) \\ 
  \bottomrule
\end{tabular}}
\caption{Intensity rate ratios (along with 95\% confidence intervals) associated with the demographic variables as well as those associated with interactions between the QIDS score and the visit profile.}
\label{democoefstable}
\end{table}

\clearpage
\subsection{Dropout model output}

\subsection{Other simulation results}

For Scenario 2, the outcome is generated from the random-intercept model $$Y_i(t)=\beta_0+\beta_1(1+t)^{-2}+\beta_2(1+t)^{-2}\log(1+t) + b_i+ \varepsilon_i(t), \quad b_i\sim N(0, \sigma_\phi^2), \quad \varepsilon_i(t) \sim N(0, \sigma_\varepsilon^2).$$ As in Scenario 1, we set $\lambda_0=1, \sigma_\phi = 1, \sigma_\varepsilon = 2$. In contrast with Scenario 1, we set $\gamma_0=0.5, \bbeta_0=(3.3,4,10.5)^\top, \tau = 3.5, c = 3$. Instead of setting positive values for $\eta_1$, we consider $-0.5$, $-1$, and $-1.5$. We adjust $\eta_0$ accordingly to achieve four informative dropout proportions: 20\%, 40\%, 60\%, and 80\%.

\begin{table}[!htbp]
\centering
\scalebox{0.95}{
\begin{tabular}{llllrrrrrrr}
  \toprule
 \% inf & Slope & Trim & Model & Bias & Emp SE & Naive SE & CP & Boot SE & Boot CP & Boot SD \\
 \midrule
20\% & 0.5 & None & IIW-NID & -1.1 & 1.8 & 1.8 & 0.90 & 1.8 & 0.90 & 0.19 \\ 
&  &  &  IIW & -1.3 & 1.8 & 1.8 & 0.88 & 1.8 & 0.87 & 0.19 \\ 
 &  &  &  IIW$\times$IPW & -0.2 & 2.8 & 2.3 & 0.94 & 2.5 & 0.96 & 1.1 \\ 
 \midrule 
20\% & 0.5 & 99.9\% &  IIW-NID & -1.1 & 1.8 & 1.8 & 0.90 & 1.8 & 0.90 & 0.19 \\ 
& &  &  IIW & -1.3 & 1.8 & 1.8 & 0.88 & 1.8 & 0.87 & 0.19 \\ 
 &  &  &  IIW$\times$IPW & -0.3 & 2.1 & 2.1 & 0.95 & 2.4 & 0.97 & 1.0 \\ 
\midrule 
20\% & 0.5& 99.5\% &  IIW-NID & -1.1 & 1.8 & 1.8 & 0.90 & 1.8 & 0.90 & 0.19 \\ 
 &  &  &  IIW & -1.3 & 1.8 & 1.8 & 0.88 & 1.8 & 0.87 & 0.19 \\ 
 &  &  &  IIW$\times$IPW & -0.5 & 1.9 & 1.9 & 0.94 & 2.0 & 0.95 & 0.36 \\ 
\midrule
20\% & 0.5 & 99\% &  IIW-NID & -1.1 & 1.8 & 1.8 & 0.90 & 1.8 & 0.90 & 0.19 \\ 
&  &  &  IIW & -1.3 & 1.8 & 1.8 & 0.88 & 1.8 & 0.87 & 0.19 \\ 
 & &  &  IIW$\times$IPW & -0.6 & 1.8 & 1.9 & 0.95 & 1.9 & 0.95 & 0.22 \\ 
   \midrule\midrule
60\% & 1.0 & None & IIW-NID & -5.0 & 2.6 & 2.6 & 0.51 & 2.6 & 0.52 & 0.35 \\ 
&  &  &  IIW & -5.3 & 2.4 & 2.4 & 0.37 & 2.4 & 0.37 & 0.31 \\ 
 &  &  &  IIW$\times$IPW & -1.3 & 8.4 & 5.0 & 0.78 & 5.6 & 0.84 & 3.0 \\ 
  \midrule 
60\% & 1.0 & 99.9\% &  IIW-NID & -5.0 & 2.6 & 2.6 & 0.51 & 2.6 & 0.52 & 0.35 \\ 
 &  &  &  IIW & -5.3 & 2.4 & 2.4 & 0.37 & 2.4 & 0.37 & 0.31 \\ 
 &  &  &  IIW$\times$IPW & -1.6 & 7.1 & 4.9 & 0.81 & 5.4 & 0.87 & 2.8 \\ 
\midrule
60\% & 1.0 & 99.5\% &  IIW-NID & -5.0 & 2.6 & 2.6 & 0.51 & 2.6 & 0.52 & 0.35 \\ 
 &  &  &  IIW & -5.3 & 2.4 & 2.4 & 0.36 & 2.4 & 0.37 & 0.31 \\ 
 &  &  &  IIW$\times$IPW & -2.6 & 4.4 & 3.9 & 0.83 & 4.6 & 0.88 & 1.7 \\ 
  \midrule
60\% & 1.0 & 99\% & IIW-NID & -5.0 & 2.6 & 2.6 & 0.51 & 2.6 & 0.52 & 0.35 \\ 
& &  &  IIW & -5.3 & 2.4 & 2.4 & 0.37 & 2.4 & 0.37 & 0.31 \\ 
 &  &  &  IIW$\times$IPW & -3.0 & 3.6 & 3.4 & 0.81 & 3.7 & 0.84 & 1.0 \\
  \midrule\midrule
80\% & 1.5 & None & IIW-NID & -8.1 & 3.8 & 3.9 & 0.44 & 4.0 & 0.46 & 0.67 \\ 
 &  &  &  IIW & -8.1 & 3.2 & 3.2 & 0.28 & 3.3 & 0.29 & 0.59 \\ 
 &  &  &  IIW$\times$IPW & -3.4 & 14 & 6.8 & 0.69 & 8.2 & 0.78 & 4.9 \\ 
   \midrule
80\% & 1.5 & 99.9\% &  IIW-NID & -8.1 & 3.8 & 3.9 & 0.44 & 4.0 & 0.46 & 0.67 \\ 
 &  &  &  IIW & -8.1 & 3.2 & 3.2 & 0.28 & 3.3 & 0.29 & 0.59 \\ 
&  &  &  IIW$\times$IPW & -3.6 & 12 & 6.8 & 0.70 & 8.1 & 0.80 & 4.6 \\ 
  \midrule
80\% & 1.5 & 99.5\% &  IIW-NID & -8.1 & 3.8 & 3.9 & 0.44 & 4.0 & 0.46 & 0.67 \\ 
 &  &  &  IIW & -8.1 & 3.2 & 3.2 & 0.28 & 3.3 & 0.29 & 0.59 \\ 
& & &  IIW$\times$IPW & -4.7 & 7.2 & 5.9 & 0.75 & 7.3 & 0.83 & 3.5 \\ 
    \midrule
80\% & 1.5 & 99\% &  IIW-NID & -8.1 & 3.8 & 3.9 & 0.44 & 4.0 & 0.46 & 0.67 \\ 
 &  &  &  IIW & -8.1 & 3.2 & 3.2 & 0.28 & 3.3 & 0.29 & 0.59 \\ 
 &  & &  IIW$\times$IPW & -5.4 & 5.7 & 5.1 & 0.73 & 5.9 & 0.78 & 2.1 \\ 
  \bottomrule
\end{tabular}}
\caption{Scenario 1: $n_\text{sim}=1000, n=200, \gamma_0=-0.336, \bbeta_0 = (16.4, -3.1)^\top, \tau=16, c=2,\lambda_0=1, \sigma_\phi=1, \sigma_\varepsilon=2$. \% inf is the percentage of subjects who were informatively censored and the slope $\eta_1$ controls ``informativeness.'' Bias is presented along with the empirical standard errors. Further, we present the naive standard errors and their corresponding coverage probabilities (CP), as well as the bootstrap standard errors along with their corresponding CPs. The standard deviation of the bootstrap standard errors is shown in the rightmost column. All three methods are compared, with four levels of trimming.}
\label{scenario1}
\end{table}

\begin{table}[!htbp]
\centering
\scalebox{0.95}{
\begin{tabular}{llllrrrrrrr}
  \toprule
 \% inf & Slope & Trim & Model & Bias & Emp SE & Naive SE & CP & Boot SE & Boot CP & Boot SD \\
 \midrule
20\% & $-0.5$ & None & IIW-NID & 0.11 & 0.38 & 0.38 & 0.92 & 0.37 & 0.93 & 0.05 \\ 
&&&  IIW & 0.12 & 0.38 & 0.37 & 0.92 & 0.37 & 0.93 & 0.05 \\ 
&&&  IIW$\times$IPW & -0.02 & 0.41 & 0.40 & 0.95 & 0.40 & 0.94 & 0.08 \\ 
\midrule
20\% & $-0.5$ & 99.9\% &  IIW-NID & 0.12 & 0.38 & 0.37 & 0.92 & 0.38 & 0.93 & 0.05 \\ 
&&&  IIW & 0.13 & 0.38 & 0.37 & 0.92 & 0.37 & 0.92 & 0.05 \\ 
&&&  IIW$\times$IPW & 0.00 & 0.39 & 0.38 & 0.95 & 0.40 & 0.95 & 0.08 \\ 
\midrule
20\% & $-0.5$ & 99.5\% &  IIW-NID & 0.13 & 0.37 & 0.37 & 0.92 & 0.37 & 0.92 & 0.04 \\ 
&&&  IIW & 0.14 & 0.37 & 0.37 & 0.92 & 0.37 & 0.92 & 0.04 \\ 
&&&  IIW$\times$IPW & 0.02 & 0.38 & 0.37 & 0.94 & 0.38 & 0.94 & 0.04 \\ 
\midrule
20\% & $-0.5$ & 99\% &  IIW-NID & 0.13 & 0.37 & 0.37 & 0.92 & 0.37 & 0.92 & 0.04 \\ 
&&&  IIW & 0.14 & 0.37 & 0.37 & 0.92 & 0.37 & 0.92 & 0.04 \\ 
&&&  IIW$\times$IPW & 0.04 & 0.38 & 0.37 & 0.95 & 0.37 & 0.94 & 0.04 \\ 
\midrule\midrule
60\% & $-1.0$ & None & IIW-NID & 0.64 & 0.42 & 0.42 & 0.64 & 0.42 & 0.65 & 0.05 \\ 
&&&  IIW & 0.69 & 0.42 & 0.41 & 0.59 & 0.41 & 0.60 & 0.05 \\ 
&&&  IIW$\times$IPW & 0.05 & 0.77 & 0.63 & 0.92 & 0.64 & 0.93 & 0.24 \\ 
\midrule
60\% & $-1.0$ & 99.9\% &  IIW-NID & 0.64 & 0.42 & 0.42 & 0.64 & 0.42 & 0.65 & 0.05 \\ 
&&&  IIW & 0.69 & 0.42 & 0.41 & 0.59 & 0.41 & 0.60 & 0.05 \\ 
&&&  IIW$\times$IPW & 0.10 & 0.66 & 0.59 & 0.93 & 0.64 & 0.95 & 0.23 \\ 
\midrule
60\% & $-1.0$ & 99.5\% &  IIW-NID & 0.64 & 0.42 & 0.42 & 0.64 & 0.42 & 0.65 & 0.05 \\ 
&&&  IIW & 0.69 & 0.42 & 0.41 & 0.59 & 0.41 & 0.60 & 0.05 \\ 
&&&  IIW$\times$IPW & 0.21 & 0.53 & 0.51 & 0.91 & 0.55 & 0.93 & 0.13 \\ 
\midrule
60\% & $-1.0$ & 99\% &  IIW-NID & 0.64 & 0.42 & 0.42 & 0.64 & 0.42 & 0.65 & 0.05 \\ 
&&&  IIW & 0.69 & 0.42 & 0.41 & 0.59 & 0.41 & 0.60 & 0.05 \\ 
&&&  IIW$\times$IPW & 0.28 & 0.47 & 0.47 & 0.89 & 0.50 & 0.91 & 0.08 \\ 
\midrule\midrule
80\% & $-1.5$ & None &IIW-NID & 1.0 & 0.55 & 0.56 & 0.54 & 0.55 & 0.53 & 0.07 \\ 
&&&  IIW & 0.95 & 0.57 & 0.57 & 0.60 & 0.57 & 0.61 & 0.08 \\ 
&&&  IIW$\times$IPW & 0.29 & 1.6 & 1.0 & 0.79 & 1.1 & 0.85 & 0.55 \\ 
\midrule
80\% & $-1.5$ & 99.9\% &  IIW-NID & 1.0 & 0.55 & 0.56 & 0.54 & 0.55 & 0.53 & 0.07 \\ 
&&&  IIW & 0.95 & 0.57 & 0.57 & 0.60 & 0.57 & 0.61 & 0.08 \\ 
&&&  IIW$\times$IPW & 0.33 & 1.4 & 1.0 & 0.81 & 1.1 & 0.87 & 0.52 \\ 
\midrule
80\% & $-1.5$ & 99.5\% &  IIW-NID & 1.0 & 0.55 & 0.56 & 0.54 & 0.55 & 0.53 & 0.07 \\ 
&&&  IIW & 0.95 & 0.57 & 0.57 & 0.60 & 0.57 & 0.61 & 0.08 \\ 
&&&  IIW$\times$IPW & 0.52 & 0.94 & 0.85 & 0.84 & 1.0 & 0.88 & 0.36 \\ 
\midrule
80\% & $-1.5$ & 99\% &  IIW-NID & 1.0 & 0.55 & 0.56 & 0.54 & 0.55 & 0.53 & 0.07 \\ 
&&&  IIW & 0.95 & 0.57 & 0.57 & 0.60 & 0.57 & 0.61 & 0.08 \\ 
&&&  IIW$\times$IPW & 0.62 & 0.76 & 0.74 & 0.82 & 0.83 & 0.87 & 0.22 \\ 
   \bottomrule
\end{tabular}}
\caption{Scenario 2: $n_\text{sim}=1000, n=200, \gamma_0=0.5, \bbeta_0 = (3.3, 4, 10.5)^\top, \tau=3.5, c=3,\lambda_0=1, \sigma_\phi=1, \sigma_\varepsilon=2$. \% inf is the percentage of subjects who were informatively censored and the slope $\eta_1$ controls ``informativeness.'' Bias is presented along with the empirical standard errors. Further, we present the naive standard errors and their corresponding coverage probabilities (CP), as well as the bootstrap standard errors along with their corresponding CPs. The standard deviation of the bootstrap standard errors is shown in the rightmost column. All three methods are compared, with four levels of trimming.}
\label{scenario2}
\end{table}

\clearpage
\subsection{Deriving the explicit form of the observed visit model}
We derive equation (\ref{fullbreakdown}) using assumption (\ref{assdN}), as well as the assumption that the competing event time is predictable.
\begin{align*}
&\E[\dd N_i(t) |\Hcal_i^O(t^-), \Xb_i(t)] \\
&= \E[\dd N_i^*(t) \zeta_i(t)| \Hcal_i^O(t^-), \Xb_i(t)]\\
&=\E[\dd N_i^*(t)| \Hcal_i^O(t^-),\Xb_i(t)]\E[\zeta_i(t)| \Hcal_i^O(t^-),\Xb_i(t)] \\
&= e^{\bgamma_0^\top\Hcal_i^O(t^-)}\dd\Lambda_0(t)\E(\1(D_i> t)\1(G_i> t)\1(L_i\geq t) | \Hcal_i^O(t^-),\Xb_i(t)) \nonumber\\
&= \1(L_i\geq t)e^{\bgamma_0^\top\Hcal_i^O(t^-)}\dd\Lambda_0(t)\E(\1(D_i> t, G_i> t) | \Hcal_i^O(t^-),\Xb_i(t)) \nonumber\\
&= \1(L_i\geq t)e^{\bgamma_0^\top\Hcal_i^O(t^-)}\dd\Lambda_0(t)\p(D_i> t,G_i> t | \Hcal_i^O(t^-), \Xb_i(t)) \nonumber\\
&= \1(L_i\geq t)e^{\bgamma_0^\top\Hcal_i^O(t^-)}\dd\Lambda_0(t)\p(D_i> t|G_i> t, \Hcal_i^O(t^-),\Xb_i(t))\p(G_i> t | \Hcal_i^O(t^-), \Xb_i(t)) \nonumber
\end{align*}

\clearpage
\subsection{Showing the estimating equation has mean zero}
We use the iterated expectation property to condition on $\Xb_i(t)$:
\begin{align*}
    \E[\bm{U}(\bbeta;\hat{\bgamma},\hat{\beeta},h)]&=\E\left\{\sum_{i=1}^n \int_0^\infty \Xb_i(t)\left\{\frac{\dd g(\mu)}{\dd \mu}\biggr|_{\mu_i(t;\bbeta)} \right\}^{-1} v(\mu_i(t;\bbeta))^{-1}h(\Xb_i(t))\right.\nonumber\\
    &\qquad \times \left.\E\left[\left\{Y_i(t)-\mu_i(t;\bbeta)\right\} \frac{\dd N_i(t)}{e^{\bgamma_0^\top \Hcal_i^O(t^-)}\p(D_i> t | G_i> t, \Hcal_i^O(t^-), \Xb_i(t))}\biggr| \Xb_i(t)\right]\right\}
\end{align*}
It remains to show that the expected value of the term in the square parentheses above is equal to zero, given our assumptions.
\begin{align*}
    & \E\left[\left\{Y_i(t)-\mu_i(t;\bbeta)\right\} \frac{\dd N_i(t)}{e^{\bgamma_0^\top \Hcal_i^O(t^-)}\p(D_i> t | G_i> t, \Hcal_i^O(t^-),\Xb_i(t))}\biggr| \Xb_i(t)\right]\\
    &= \E\left[\left\{Y_i(t)-\mu_i(t;\bbeta)\right\} \frac{\E[\dd N_i(t)| \Hcal_i^O(t^-), \Xb_i(t) ]}{e^{\bgamma_0^\top \Hcal_i^O(t^-)}\p(D_i> t | G_i> t,\Hcal_i^O(t^-),\Xb_i(t))}\biggr| \Xb_i(t)\right]\\
    &= \E\left[\left\{Y_i(t)-\mu_i(t;\bbeta)\right\}\1(L_i\geq t) \right.\\
    &\qquad \times \left.\frac{e^{\bgamma_0^\top\Hcal_i^O(t^-)}\dd\Lambda_0(t)\p(D_i>t|G_i> t, \Hcal_i^O(t^-), \Xb_i(t))\p(G_i> t|\mathcal{H}_i^O(t^-),\Xb_i(t))}{e^{\bgamma_0^\top \Hcal_i^O(t^-)}\p(D_i> t | G_i> t, \Hcal_i^O(t^-),\Xb_i(t))}\biggr| \Xb_i(t)\right]\\
    &= \E\left[\left\{Y_i(t)-\mu_i(t;\bbeta)\right\} \dd \Lambda_0(t)\1(L_i\geq t)\p(G_i> t|\mathcal{H}_i^O(t^-),\Xb_i(t))\biggr| \Xb_i(t)\right]\\
    &= \E\left[\left\{Y_i(t)-\mu_i(t;\bbeta)\right\} \dd \Lambda_0(t)\1(L_i\geq t)\E(\1(G_i> t)|\mathcal{H}_i^O(t^-),\Xb_i(t))\biggr| \Xb_i(t)\right]\\
    &=\E\left[\left\{Y_i(t)-\mu_i(t;\bbeta)\right\} \dd \Lambda_0(t)\1(L_i\geq t)\1(G_i>t)\biggr| \Xb_i(t)\right]\\
    &= \dd \Lambda_0(t)\E[\1(G_i>t)|\Xb_i(t)]\E\left[\E\left[\left\{Y_i(t)-\mu_i(t;\bbeta)\right\} \1(L_i\geq t)\biggr| \Xb_i(t),\bar\xi_i^L(t)=0 \right]\biggr| \Xb_i(t)\right]\\
    &= \dd \Lambda_0(t)\p(G_i>t|\Xb_i(t))\E\left[\1(L_i\geq t)\E\left[\left\{Y_i(t)-\mu_i(t;\bbeta)\right\} \biggr| \Xb_i(t),\bar\xi_i^L(t)=0 \right]\biggr| \Xb_i(t)\right]=0
\end{align*}
since the innermost expectation is equal to zero by assumption.

\end{document}